\documentclass[12pt]{article}
\pdfoutput=1

\usepackage[utf8]{inputenc}
\usepackage[left=2.55cm, right=2.55cm, top=2.55cm, bottom=2.55cm]{geometry}
\usepackage{amsmath,amssymb,amsbsy,mathtools}
\usepackage{slashed}
\usepackage{xcolor}
\usepackage{graphicx}
\usepackage{url}
\usepackage{cancel}
\usepackage{cite}
\usepackage[colorlinks=true,allcolors=blue,pdfborder={0 0 0},linktocpage=false,pdfencoding=auto]{hyperref}
\usepackage{tabularx,booktabs}
\usepackage{multicol}
\usepackage{multirow}
\usepackage{feynmp}
\usepackage{units}
\usepackage{xspace}
\usepackage[labelfont=bf]{caption}
\usepackage[section]{placeins}
\usepackage{subcaption}
\usepackage{soul}
\usepackage{bbold}
\usepackage{parskip}
\usepackage{float}
\usepackage{tabulary}
\usepackage{empheq}
\usepackage{tabu}

\definecolor{darkred}{rgb}{0.6,0,0}
\definecolor{darkpurple}{rgb}{0.5,0,0.5}

 \newcommand{\code}[1]{\texttt{#1}}
\newcommand{\beqn}{\begin{eqnarray}}
\newcommand{\eeqn}{\end{eqnarray}}

\begin{document}
\author{Amin Aboubrahim$^{a,b}$\footnote{\href{mailto:abouibrah@hartford.edu}{abouibrah@hartford.edu}}\,\,,~~Andrew H. Giman$^b$\footnote{\href{mailto:andrewgiman1@gmail.com}{andrewgiman1@gmail.com}}~~and Pran Nath$^c$\footnote{\href{mailto:p.nath@northeastern.edu}{p.nath@northeastern.edu}}
 \\~\\
$^{a}$\textit{\normalsize Department of Physics, University of Hartford,} \\
\textit{\normalsize 200 Bloomfield Ave., West Hartford, CT 06117, U.S.A.} \\
$^{b}$\textit{\normalsize Department of Physics and Astronomy, Union College,} \\
\textit{\normalsize 807 Union Street, Schenectady, NY 12308, U.S.A.} \\
$^{c}$\textit{\normalsize Department of Physics, Northeastern University,} \\
\textit{\normalsize 111 Forsyth Street, Boston, MA 02115-5000, U.S.A.} \\
}

\title{\vspace{-2cm}
\vspace{1cm}
\large \bf
Cosmology of axion dark energy in supersymmetric models and constraints on high scale parameters
 \vspace{0.5cm}}
\date{}
\maketitle

\vspace{0.5cm}

\begin{abstract}

An analysis is given of interacting dark energy and dark matter where the dark energy
is assumed to be an ultralight axionic field with a pseudo-Nambu-Goldstone Boson potential which is in general a superposition of $N$ number of  cosine terms  motivated by supergravity and string models with a $U(1)$ global symmetry, where the symmetry is broken by instanton effects. The case $N=2$ is investigated in detail and a fit to cosmological data is performed where it is found that a better fit is obtained in comparison with the $N=1$ case. The fits also constrain high scale parameters, i.e., the axion decay constant which is determined to be sub-Planckian, a result consistent with string theory that disfavors trans-Planckian axion decay constant. Furthermore, the dark energy-dark matter interaction strength is constrained to be feeble, i.e., $\lambda\lesssim 4\times 10^{-6}$ m$_{\rm Pl}^{-2}$ Mpc$^{-2}$. We study possible implications of this type of potential on the Hubble tension and on the dynamics of the dark energy equation of state using the DESI-DR2 data. For the cases $N=3,4$, the analysis exhibits the phenomenon of transmutation even in the absence of coupling to dark matter, where thawing quintessence transmutes to freezing quintessence. The analysis is internally consistent in its treatment of the dark energy-dark matter interaction as it is based on an underlying Lagrangian, in contrast with several previous works where the sources are chosen in an ad hoc manner to satisfy energy conservation. 

\end{abstract}

\numberwithin{equation}{section}

\newpage

{  \hrule height 0.4mm \hypersetup{linktocpage=true} \tableofcontents
\vspace{0.5cm}
 \hrule height 0.4mm}

\section{Introduction}

A compelling alternative to the cosmological constant $\Lambda$ as an explanation for dark energy (DE) is quintessence~\cite{Caldwell:1997ii,Ratra:1987rm}.
This class of models represents a slowly rolling scalar field $\phi$ whose energy density evolves in time and drives the late-time cosmic acceleration. Furthermore, quintessential dark energy provides a scenario with a dynamical equation of state, unlike $\Lambda$ whose equation of state (EoS) is $w_\Lambda=-1$ throughout cosmic history. This dynamical behavior of the equation of state $w$ allows a classification of quintessence into thawing and freezing models. In thawing quintessence~\cite{Scherrer:2007pu,Caldwell:2005tm,Wolf:2023uno,Wolf:2024eph}, the Hubble friction dominates in the early universe and the field remains frozen so that $w_\phi\simeq -1$. When the mass of the field drops below the Hubble parameter at late times, $w_\phi$ starts to deviate away from $-1$. In freezing models, $w_\phi$ dips towards $-1$ at late times because of the shallow nature of the potential which, for a scaling freezing model~\cite{Ferreira:1997hj,Copeland:1997et}, is taken to be a double exponential potential. Other than the DE equation of state, quintessence can imprint distinctive features on structure formation on all scales, especially during the non-linear regime~\cite{Alimi:2009zk}. 

An ultralight axion as a pseudo-Nambu-Goldstone boson (pNGB)~\cite{Frieman:1995pm} also belongs to the class of quintessence models with a potential of the form
\begin{equation}
    V(\phi)=\mu^4\Big(1+\cos(\phi/F)\Big),
\end{equation}
where $\mu$ and $F$ are constants and $\phi$ is the axionic field. It is expected that $F\sim M_{\rm Pl}$ (Planck mass) and $\mu^4\sim \rho_{\rm cr}$ is today's critical density of the universe.  This potential has been investigated thoroughly in the literature~\cite{Urena-Lopez:2025rad,Lee:2025yvn,Abreu:2025zng,Shajib:2025tpd,Berbig:2024aee,Bhattacharya:2024kxp,Choi:2021aze,Smer-Barreto:2015pla,Dutta:2006cf,Caldwell:2005tm,Hall:2005xb,Kawasaki:2001bq,Waga:2000ay,Ng:2000di,Viana:1997mt,Frieman:1997xf,Coble:1996te,Lin:2025gne} and tested as a possibility for the resolution of the Hubble tension, which refers to a discrepancy between early-time inference of the Hubble parameter from measurements of the cosmic microwave background and local measurements of the Hubble using supernova data~\cite{Poulin:2023lkg,Cai:2023sli,Efstathiou:2024dvn,CosmoVerseNetwork:2025alb,Abdalla:2022yfr}. Recent results from the SH0ES collaboration~\cite{Riess:2021jrx,Breuval:2024lsv} now stands at a $5\sigma$ deviation from the Planck collaboration's result~\cite{Planck:2018vyg}. It was shown that simple uncoupled quintessence models are not successful at resolving or ameliorating the Hubble tension since they tend to produce $w>-1$ which lowers the expansion rate at fixed late‐time cosmic distances and thereby tends to worsen the discrepancy between local and CMB-inferred Hubble parameter~\cite{Sabla:2021nfy,Banerjee:2020xcn,Lee:2022cyh}, whereas some coupled quintessence can have a mild relief~\cite{Aboubrahim:2025qat,Aboubrahim:2024cyk,Aboubrahim:2024spa,SanchezLopez:2025uzw}. Furthermore, in light of the latest Dark Energy Spectroscopic Instrument (DESI) DR2 BAO and redshift‐space distortion measurements, there is growing evidence for an evolving equation of state with indication of phantom crossing $w<-1$~\cite{DESI:2024mwx,DESI:2025zgx}\footnote{The apparent crossing of the phantom divide as shown by DESI does not necessarily lead to violation of the Null Energy Condition~\cite{Wolf:2024eph,Ramadan:2024kmn,Cortes:2024lgw}.}, an interpretation based on the Chevallier-Polarski-Linder (CPL) parameterization~\cite{Chevallier:2000qy,Linder:2002et} of the EoS, $w(a)=w_0+(1-a)w_a$. Thus, DESI DR2 reports a preference for $w_0>-1$ and $w_a<0$ with the $\mathrm{DESI~BAO}+\mathrm{CMB}+\mathrm{DESY}5$ data set at the $4.2\sigma$ level. Quintessence models produce a rather more complicated EoS parametrization~\cite{Aboubrahim:2025qat,Tsujikawa:2013fta} and do not necessarily lead to phantom crossing. The analysis of ref.~\cite{Shlivko:2024llw} shows that some quintessence models can sometimes lead to crossing of the phantom divide, but this apparent issue can simply be a byproduct of extrapolating the CPL parametrization beyond the region of validity where data fitting is performed.

An axion-like pNGB quintessence is one of the most theoretically motivated classes of dynamical dark energy models, owing to shift‐symmetry protection and potential connection to UV physics (supergravity and strings)~\cite{Berbig:2024aee}. Many inflationary physics models use a similar field that slowly rolls along its potential and some works have attempted to develop a unified framework of inflation and expansion driven by dark energy~\cite{Cicoli:2024yqh,Cicoli:2021skd}. In this work, we focus on an axionic potential composed of a superposition of cosine terms~\cite{Svrcek:2006yi,Halverson:2017deq,Nath:2017ihp}, a feature that is ubiquitous in many supergravity and string models. This dark energy potential is embedded in a field-theoretic model of cosmology with an interaction term between dark matter (DM) and dark energy. We investigate the effect of this potential on the Hubble tension and the dark energy EoS.

\section{Axion potential in an axion landscape}

In a variety of string and supergravity models, the axionic potential in general involves superposition of cosine 
terms of the form~\cite{Svrcek:2006yi,Halverson:2017deq,Nath:2017ihp}
\begin{align}
V(a) =\sum_{k=1}^N  \mu_k^4\cos\left(\frac{ k a}{F}\right).
\label{witten}
\end{align}
Such potentials arise naturally in landscape models involving multiple axions and anomalous global 
$U(1)_{X_n}$ symmetries which are broken at lower scales by instanton effects. Here we will consider the simplest
possibility of one global $U(1)_X$ symmetry, but $N$ number of axionic fields carrying the same $U(1)$ quantum number. Thus, suppose  we have a set of fields $\Phi_i$ ($i=1, \cdots, N$) where $\Phi_i$ carry the same charge $Q$ under the shift symmetry and the fields $\bar \Phi_i$ ($i=1, \cdots, m$) carry the opposite charge $-Q$. Then under $U(1)_X$ transformations one has
\begin{align}
\Phi_i\to e^{i Q \lambda} \Phi_i, \quad\quad \bar \Phi_i\to e^{-i Q \lambda} \bar \Phi_i, \quad (i=1, \cdots, m)\,.
\end{align}
The superfields ${\Phi}_{i}$ have an expansion ${\Phi}_{i} = {\phi}_{i} + \theta  {\chi}_{i} + \theta  \theta  {F}_{i}\,,$ where ${\phi}_{i}$ is a complex scalar field consisting of the saxion (the real part) and the axion (the imaginary
part), ${\chi}_{i}$ is the axino, and ${F}_{i}$ is an auxiliary field.
Similarly, the superfields $\bar {\Phi}_{i}$ have an expansion:
$\bar {\Phi}_{i} = \bar {\phi}_{i} + \bar \theta  \bar {\xi}_{i} + \bar\theta \bar  \theta  \bar{F}_{i}$.
We consider now a superpotential of the form 
\begin{align}
W = W_s(\Phi,\bar \Phi) + W_ {sb} (\Phi, \bar \Phi)\,,
\label{wsn}
\end{align}
where $W_s$ is the part that depends on the fields $\Phi_i,  \bar \Phi_i$ and is invariant under the shift symmetry.
$W_{sb}$ is a  part which breaks the shift symmetry and has the form
\begin{align}
W_{sb} = \sum_{i} C_i(\Phi, \bar \Phi) e^{-T_i}\,,
\end{align}
 where $T_i$ is the action of the $i$-th instanton. In general, gauge invariance and holomorphy allow non-perturbative
 terms of the type
 \begin{align}
  C \frac{\Phi^n}{M_{P}^{n-3}} e^{-T}, \quad ~~\bar C \frac{\bar \Phi^n}{M_{P}^{n-3}} e^{-T}\,,
 \end{align}
where the detailed structure will depend on the instanton zero modes.
Thus, we assume the following form for $W(\Phi, \bar \Phi)$
\begin{equation}
W \left( \Phi,\bar \Phi \right) = \sum_{k = 1}^{m} \sum_{l = 1}^{m} \left( {\mu}_{k  l} {\Phi}_{k} {\bar \Phi}_{l} + \frac{{\lambda}_{k  l}}{2 M} {\left( {\Phi}_{k} {\bar \Phi}_{l} \right)}^{2} \right) + \sum_{k = 1}^{m} \sum_{l = 1}^{q} {C}_{k  l} {\Phi}_{k}^{l} + \sum_{k = 1}^{m} \sum_{l = 1}^{q} {\bar{C}}_{k  l} {\bar{\Phi}}_{k}^{l}\,.
\label{wsn1}
\end{equation}
Here the terms in the first brace on the right hand side are invariant under the shift symmetry  
while the remaining terms on the right hand side violate the shift symmetry.
 The variation of the superpotential
 with respect to $\Phi_k$ and $\bar \Phi_k$ generates  the constraints that determine the vacuum expectation values (VEVs) of $\Phi_k$ and 
 $\bar\Phi_k$. We assume CP conserving vacua so that the VEVs of the CP odd axionic fields vanish while we set
 $f_k=\langle\Phi_k\rangle$ and $\bar f_k= \langle\bar \Phi_k\rangle$. The constraint equations arising from the variation of the superpotential with respect to $\Phi_k$ and $\bar \Phi_k$ are  
\begin{equation}
\begin{aligned}
\frac{\partial W \left( \phi , \overline{\phi} \right)}{\partial {\phi}_{k}} = 0\,, \quad ~~
\frac{\partial W \left( \phi , \overline{\phi} \right)}{\partial {\overline{\phi}}_{k}} = 0\,.
\label{saxion-1}
\end{aligned}
\end{equation}
We may parametrize $\phi_k$ and $\bar \phi_k$ so that 
\begin{align}
\phi_k = (f_k + \rho_k) e^{ia_k/F_k}, \quad ~~\bar\phi_k = (\bar f_k + \bar \rho_k) e^{i\bar a_k/\bar F_k}\,,
\end{align}
where $f_k= \langle\phi_k\rangle ,~\bar f_k= \langle\bar\phi_k\rangle$ and  $(\rho_k, a_k)$ and $(\bar \rho_k, \bar a_k)$ 
are the fluctuations of the quantum fields around their vacuum expectation values  $f_k~(\bar f_k)$. 
They are constrained by the stability conditions for the saxions, Eqs. (\ref{saxion-1}), which give 
\begin{equation}
\begin{aligned}
\sum_{l = 1}^{m} \left( {\mu}_{k  l} {\bar{f}}_{l} + \frac{{\lambda}_{k  l}}{M} {f}_{k} {\bar{f}}_{l}^{2} \right) + \sum_{l = 1}^{q} l  {C}_{k  l} {f}_{k}^{l - 1} = 0\,,\\
\sum_{l = 1}^{m} \left( {\mu}_{l  k} {f}_{l} + \frac{{\lambda}_{l  k}}{M} {f}_{l}^{2} {\bar{f}}_{k} \right) + \sum_{l = 1}^{q} l  {\bar{C}}_{k  l} {\bar{f}}_{k}^{l - 1} = 0\,.
\label{ssb}
\end{aligned}
\end{equation}
We focus here on the scalar potential for the axions and thus we expand around the minimum of the potential
of the saxions, i.e., we
set $\rho_k=0=\bar \rho_k$.
Using the saxion stability conditions given by Eq.~(\ref{ssb}) a somewhat lengthy computation gives
the scalar potential. 
There are $2m$ number of axionic fields $a_1, \cdots, a_m$ and $\bar a_1, \cdots, \bar a_m$.
Since there is only one $U(1)$ shift symmetry, 
we can pick a basis where 
only one linear combination of it is variant under the shift symmetry and all others are 
invariant. We label this new basis $a_-, a_+, \tilde a_1, \tilde a_2, \cdots, \tilde a_{m-1};  \bar {\tilde b}_1,$ 
$\bar {\tilde b}_2,$ $\cdots,$ $\bar  {\tilde b}_{m-1}$. Here,  
 only $a_-$ is sensitive to the shift symmetry while all the remaining fields are invariant. The field $a\equiv a_-$
 is the axionic field of interest and its potential can be extracted from the rest and has the following form~\cite{Nath:2017ihp}   
\begin{align}
{V} (a) =& \sum_{k=1}^N c_{k} \left[1+ \cos\left(\frac{k a}{F}\right)\right] + 
\sum_{\ell=1}^N\sum_{r=\ell+1}^N 
c_{rl} \left[1+ \cos\left(\frac{r-l}{F} a\right) \right], 
\label{twosums}
\end{align}
where the coefficients $c_k$ and $c_{rl}$ are given by
\begin{align}
 c_k =&  2 \sum_{r = 1}^{N} r  \left( {C}_{k  r} {f}_{k}^{r - 1}  \sum_{l = 1}^{N} l  {A}_{k  l} {f}_{k}^{l - 1} + {\overline{C}}_{k  r}  {\overline{f}}_{k}^{r - 1} \sum_{l = 1}^{N} l \, {\overline{C}}_{k  l} {\overline{f}}_{k}^{l - 1} \right),  \nonumber\\
c_{rl}=& - 2 l \sum_{k=1}^N r  \left( {C}_{k  l} {C}_{k  r} {f}_{k}^{l + r - 2} + {\overline{C}}_{k  l} {\overline{C}}_{k  r} {\overline{f}}_{k}^{l + r - 2} \right).
\label{slow1}
\end{align}
Here, one finds that only the effective axion decay constant appears which is given by
\begin{align}
F= \sqrt{\sum_{k = 1}^{m} {f}_{k}^{2} + \sum_{k = 1}^{m} {\overline{f}}_{k}^{2} }\,.
\label{fe}
\end{align}
In the analysis of this work we will neglect the double sum terms in the potential Eq.~(\ref{twosums}) 
involving the coefficients $c_{rl}$ and use the expansion terms in the single sum with coefficients $c_{k}$.
A remarkable aspect of Eq.~(\ref{slow1}) is that the cosine functions depend only on an effective decay constant $F$.

\section{The model}

We consider a model of quintessence dark energy interacting with dark matter~\cite{Aboubrahim:2025qat,Aboubrahim:2024cyk,Aboubrahim:2024spa}, with two spin zero fields $\chi$ and $\phi$,  where the former represents dark matter and the latter is quintessence. This model, dubbed QCDM, has an action that is given by
\begin{align}
 &S_{\rm QCDM}=\int \text{d}^4 x\,\sqrt{-g}
  \left[\frac{1}{16\pi G}R+ {\cal L}_{\rm QCDM}\right],
\end{align}
with a QCDM Lagrangian written as
\begin{align}
 & {\cal L}_{\rm QCDM}= \frac{1}{2}\partial_\mu \phi\partial^\mu\phi+\frac{1}{2}\partial_\mu \chi\partial^\mu\chi-
  V(\phi,\chi)\,.
\end{align}
The different terms in the total potential $V(\phi,\chi)$ are
\begin{align}
  &V(\phi,\chi)=V_1(\chi)+V_2(\phi)+ V_{\rm int}(\chi,\phi),
\label{n2}    
\end{align}
where we take $V_1(\chi)$ to be a simple mass term for DM and $V_2(\phi)$ is an axion-like potential
\begin{align}
\label{v1}
&V_1(\chi)=\frac{1}{2}m_\chi^2\chi^2, \\
\label{v2}
&V_2(\phi)=\mu^4\sum_{n=1}^Nc_n\left[1+\cos\left(\frac{n\,\phi}{F}\right)\right]\,.
\end{align}
The quintessence potential, Eq.~(\ref{v2}), is the first term on the right-hand-side of Eq.~(\ref{twosums}) which represents a superposition of harmonic terms in the axionic potential. As we have shown in the previous section, such potentials do arise in high scale models such as in SUGRA\cite{Nath:2017ihp} and 
strings~\cite{Panda:2010uq}. The interaction term $V_{\rm int}(\chi,\phi)$ is a particle-physics-inspired interaction potential between the fields $\phi$ and $\chi$ and is given by
\begin{align}
&V_{\rm int}(\phi,\chi)=\frac{\lambda}{2}\chi^2\phi^2.
\label{v3}
\end{align}
The total energy density of the DM and DE fields is defined by
\begin{equation}
\rho=\rho_\phi+\rho_\chi-V_{\rm int}(\chi,\phi)\,.
\label{n7}
\end{equation}
We note that $\rho_\phi$ and $\rho_\chi$ each contain $V_{\rm int}$ which requires 
to remove on factor of $V_{\rm int}$ on the right hand side so that $\rho$ is properly defined to be the total density. Thus we write 
\begin{align}
\label{n8a}
\Omega_{0i}&=\frac{\rho_i}{\rho_{0,\rm crit}}(1-\epsilon_i), 
   ~~~~~~\sum_i \Omega_{0i}=1,~~~~~~
\epsilon_i=\frac{V_{\rm int}}{\sum_i \rho_i}\,,
\end{align}
where $\rho_{0,\rm crit}$ is the critical energy density of the universe today.

To our knowledge, a superposition in an axionic potential for a single field representing dark energy and its effect on cosmology has not been studied before. However, a superposition of different axions has been considered in the literature~\cite{Muursepp:2024kcg,Muursepp:2024mbb,Gangopadhyay:2022dbm,Ettengruber:2023tac}. The interacting model we present here, equipped with an axion superposition, provides a link between field-theoretic high scale models and cosmological observations from currently available data. In the next sections, we present the background and perturbation evolution equations based on the QCDM model. 

We pause here to highlight the fact that our analysis is internally consistent in the treatment of the interaction of dark energy with dark matter in contrast to several previous works. Thus, previous analyses have frequently used a two-fluid model involving dark energy ($\phi$)  and dark matter ($\chi$) whose continuity equations read
 \begin{align*}
&\rho^\prime_\phi+3\mathcal{H}(1+w_\phi)\rho_\phi=Q\,, \\
&\rho^\prime_\chi+3\mathcal{H}(1+w_\chi)\rho_\chi=-Q\,,
\end{align*}
where $Q$ represents the interaction  between the two fields.  
In two previous papers~\cite{Aboubrahim:2024spa,Aboubrahim:2024cyk} we have shown  that the fluid  equations  being used
as exhibited above are inconsistent for any non-vanishing $Q$,  since they imply that the field $\phi$ is determined in terms of $\chi$ which is phenomenologically incorrect. Further, we 
have shown that the fluid equations above cannot arise from a Lagrangian theory, and the consistency of the conservation of the stress-energy tensor is imposed by hand which is in contradiction with  diffeomorphism of the theory. Unfortunately, this error has crept in the cosmology literature for the past quarter century~\cite{Abdalla:2022yfr} and has gone unnoticed. As noted above, this issue was discussed in depth in  refs.~\cite{Aboubrahim:2024spa,Aboubrahim:2024cyk}.

Another point we wish to emphasize is that  it is generally agreed that our ultimate goal is to achieve a UV complete theory. In this regard the supergravity and string models are UV complete. Within this context we have constructed a supergravity model incorporating shift symmetry which leads to a potential involving several cosine terms. These arise naturally within a consistent supergravity theory which is Lagrangian-based, indicating the robustness of the model. Further, while there are a large number of works analyzing DESI's data, not all analyses are internally consistent, specifically those using the two-fluid model as noted above.

\section{Evolution of the background and perturbation fields}

We consider a flat, homogeneous and isotropic universe characterized by the Friedmann-Lema\^{i}tre-Robertson-Walker (FLRW) metric. The line element is
\begin{equation}
    \text{d}s^2=a^2(-\text{d}\tau^2+\gamma_{ij}\text{d}x^i \text{d}x^j),
\end{equation}
where $a$ is the time-dependent scale factor, $\gamma_{ij}$ are the spatial components of the metric and $\tau$ is the conformal time which is related to the cosmic time $t$ by $\text{d}\tau=\text{d}t/a(t)$. The dark matter and dark energy fields contribute to the stress-energy tensor, and so their energy densities and pressure can be inferred from it. Thus, the $00$ component of the stress-energy tensor gives the energy densities, while the $ij$ component gives the pressure. As a result, the energy density and pressure of the field $\chi$ are
\begin{align}
    \rho_\chi&=\frac{1}{2a^2}\chi^{\prime \,2}+V_1(\chi)+V_{\rm int}(\chi,\phi), \\
    p_\chi&=\frac{1}{2a^2}\chi^{\prime \,2}-V_1(\chi)-V_{\rm int}(\chi,\phi)\,,
\end{align}
while that of $\phi$ are 
\begin{align}
    \rho_\phi&=\frac{1}{2a^2}\phi^{\prime \,2}+V_2(\phi)+V_{\rm int}(\chi,\phi), \\
    p_\phi&=\frac{1}{2a^2}\phi^{\prime \,2}-V_2(\phi)-V_{\rm int}(\chi,\phi)\,.
\end{align}
Note that $^\prime$ represents a derivative with respect to conformal time.
In order to determine the energy density and pressure, we need to track the evolution of the two background fields $\chi$ and $\phi$ by solving the Klein-Gordon (KG) equations 
\begin{align}
\label{kgc0}
&\chi^{\prime\prime}+2\mathcal{H}\chi^\prime+a^2(V_{1,\chi}+V_{\text{int},\chi})=0, \\
&\phi^{\prime\prime}+2\mathcal{H}\phi^\prime+a^2(V_{2,\phi}+V_{\text{int},\phi})=0\,,
\label{kgp0}
\end{align}
where $\mathcal{H}=a^\prime/a$ is the conformal Hubble parameter and $V_{,X}\equiv \partial V/\partial X$. 
It is well known that a DM potential of the form given by Eq.~(\ref{v1}), poses a difficulty in numerically solving the KG equation, Eq.~(\ref{kgc0}). This difficulty arises when $\mathcal{H}/m_\chi\ll 1$~\cite{Turner:1983he,Urena-Lopez:2015gur}, resulting in rapid oscillations of the field $\chi$. Even though these oscillations make a numerical solution intractable, they are important in the context of giving $\chi$ the character of a dark matter component. When oscillations ensue, the energy density of $\chi$ redshifts as matter, i.e., $\rho_\chi\sim a^{-3}$. But to circumvent this numerically difficulty, we will rewrite the KG equation, Eq.~(\ref{kgc0}), as a set of three coupled first order differential equations in some new variables. 

To simplify the analytical calculations, we define the DM energy density without the interaction term, i.e., $\tilde{\rho}_\chi=\rho_\chi-V_{\rm int}$. Then, the modified energy density fraction becomes
\begin{align}
    \tilde{\Omega}_\chi\equiv\frac{\tilde{\rho}_\chi}{\rho_{\rm cr}}&=\frac{\kappa^2 a^2}{3\mathcal{H}^2}\Big(\frac{1}{2a^2}\chi^{\prime 2}+V_1(\chi)\Big)=\frac{\kappa^2 a^2}{3\mathcal{H}^2}\tilde{\rho}_\chi\,,
    \label{om-tilde}
\end{align}
where $\kappa\equiv\sqrt{8\pi G}$. Based on Eq.~(\ref{om-tilde}), and alongside $\tilde\Omega_\chi$, we introduce two new dimensionless variables, $\theta$ and $y$, defined as~\cite{Copeland:1997et,Garcia-Arroyo:2024tqq}
\begin{align}
    \label{var-tilde-1}
    &\tilde{\Omega}_\chi^{1/2}\sin\left(\frac{\theta}{2}\right)=\frac{\kappa \chi^\prime}{\sqrt{6}\mathcal{H}}, \\
    \label{var-tilde-2}
    &\tilde{\Omega}_\chi^{1/2}\cos\left(\frac{\theta}{2}\right)=\frac{\kappa a V^{1/2}_{1}}{\sqrt{3}\mathcal{H}}, \\
    &y=-\frac{2\sqrt{2}\,a}{\mathcal{H}}\partial_\chi V^{1/2}_{1}.
    \label{var-tilde-3}
\end{align}
The rapid oscillations of the DM field $\chi$ now appear in the new variable $\theta$. One can then apply to $\theta$ a cut-off scale that numerically tames these oscillations. One can show that the DM equation of state (EoS), $w_\chi=p_\chi/\rho_\chi$, can be written solely in terms of $\theta$. In fact, it is simply $w_\chi=-\cos\theta$. Thus, the rapid oscillations of the field appear in the DM EoS, where $w_\chi$ oscillates between $-1$ and $+1$, resulting in an average of $\langle w_\chi\rangle =0$, which is exactly what a DM component behaves like. So, the cut-off scale that is imposed on $\theta$ when oscillations begin ensures that $w_\chi=0$, and that the numerical integration can continue without impediment.  

After a lengthy calculation, we arrive at the differential equations governing the evolution of the three dimensionless parameters
\begin{align}
\label{bk1}
&\Omega^\prime_\chi=3\mathcal{H}\Omega_\chi(w_T-w_\chi)+\frac{\kappa^2 a^2}{3\mathcal{H}^2}\phi^\prime V_{\text{int},\phi}\,, \\
\label{bk2}
&\theta^\prime=-3\mathcal{H}\frac{\Omega_\chi}{\tilde{\Omega}_\chi}\sin\theta+\mathcal{H}y-\sqrt{\frac{2}{3}}\frac{\kappa\, a^2}{\mathcal{H}\tilde{\Omega}_\chi^{1/2}} V_{\text{int},\chi}\cos\frac{\theta}{2}\,, \\
\label{bk3}
&y^\prime=\frac{3}{2}{\cal H}(1+w_T)y\,,
\end{align}
where $w_T=\sum p_i/\sum\rho_i$ is the total EoS, with the sum running over all species (baryons, photons, neutrinos, DM and DE), and
\begin{equation}
    \tilde\Omega_\chi=\Omega_\chi-\frac{\kappa^2a^2}{3\mathcal{H}^2}V_{\rm int}\,.
\end{equation}
We know that the universe we live in is not homogeneous. The growth of perturbations in the early universe played a vital role in shaping the universe we see today. To determine these perturbations in the DM and DE fields, we first define the perturbed line element in the synchronous gauge as $\text{d}s^2=a^2(\tau)[-\text{d}\tau^2 +(\delta_{ij}+h_{ij})\text{d}x^i \text{d}x^j]$, where $h_{ij}$ is a metric perturbation. We then perturb the fields so that, to first order, we have: $\phi(\Vec{x},t)\simeq \phi_0(t)+\phi_1(\Vec{x},t)$ and $\chi(\Vec{x},t)\simeq \chi_0(t)+\chi_1(\Vec{x},t)$. 

Next, we perturb the stress-energy tensor, so that $T^{\mu}_{\nu}=\bar{T}^{\mu}_{\nu}+\delta T^{\mu}_{\nu}$, with
\begin{align}
    T^{0}_0&=-\rho-\delta\rho\,, \nonumber \\
    T^{0}_i&=(\rho+p)(v_i-B_i)\,, \nonumber \\
    T^{i}_0&=-(\rho+p)v_i\,, \nonumber \\
    T^{i}_j&= (p+\delta p)\delta^i_j+p\Pi^i_j \,.
\end{align}
Here $\Pi^i_j$ represents the anisotropic stress, $v_i$ the 3-velocity and  $\delta\rho$ and $\delta p$ are the density and pressure perturbations, respectively. It immediately follows that the density and pressure perturbations of the two fields, $\chi$ and $\phi$, in the synchronous gauge are given by
\begin{align}
    \label{drhop}
    \delta\rho_\phi&=\frac{1}{a^2}\phi_0^\prime\phi_1^\prime+({V}_2+{V}_{\rm int})_{,\phi}\phi_1+{V}_{\text{int},\chi}\chi_1, \\
    \delta p_\phi&=\frac{1}{a^2}\phi_0^\prime\phi_1^\prime-({V}_2+{V}_{\rm int})_{,\phi}\phi_1-{V}_{\text{int},\chi}\chi_1, \\
    \delta\rho_\chi&=\frac{1}{a^2}\chi_0^\prime\chi_1^\prime+({V}_1+{V}_{\rm int})_{,\chi}\chi_1+{V}_{\text{int},\phi}\phi_1, \\
    \delta p_\chi&=\frac{1}{a^2}\chi_0^\prime\chi_1^\prime-({V}_1+{V}_{\rm int})_{,\chi}\chi_1-{V}_{\text{int},\phi}\phi_1.
    \label{dpchi}
\end{align} 
From the perturbed stress-energy tensor, we have the off-diagonal term $\delta T^0_i=-a^{-2}\phi_0^\prime \delta\phi_{,i}$. Taking the spatial derivative and switching to momentum space, we obtain the velocity divergence $\Theta=ik^i v_i$ of the fields
\begin{align}
    (\rho_\phi+p_\phi)\Theta_\phi&=\frac{k^2}{a^2}\phi_0^\prime\phi_1, \\
    \label{divchi}
    (\rho_\chi+p_\chi)\Theta_\chi&=\frac{k^2}{a^2}\chi_0^\prime\chi_1\,.
\end{align}
Determining the density and pressure perturbations of the fields requires evaluating the perturbations $\chi_1$ and $\phi_1$ (as well as the background fields).  This is achieved by solving the Klein-Gordon equations of the two fields
\begin{align}
\label{KGp1}
&\phi_1^{\prime\prime}+2\mathcal{H}\phi_1^\prime+(k^2+a^2 V_{,\phi\phi})\phi_1+a^2{V}_{,\phi\chi}\chi_1+\frac{1}{2}h^\prime\phi_0^\prime=0, \\
&\chi_1^{\prime\prime}+2\mathcal{H}\chi_1^\prime+(k^2+a^2{V}_{,\chi\chi})\chi_1+a^2{V}_{,\chi\phi}\phi_1+\frac{1}{2}h^\prime\chi_0^\prime=0,
\end{align}
where $V_{,XX}=\partial^2 V/\partial X^2$ and $V_{,XY}=\partial^2 V/\partial X\partial Y$.
The same numerical difficulty we faced in solving the KG equation of the background field $\chi$ also arises for the perturbation $\chi_1$. To mitigate this, we proceed in a similar fashion as we did for the background equations. First, we define the DM density perturbations by absorbing the interaction term into the $\delta\tilde{\rho}_\chi$, so that
\begin{equation}
   \delta\tilde{\rho}_\chi=\delta\rho_\chi-V_{\text{int},\phi}\phi_1-V_{\text{int},\chi}\chi_1=\frac{1}{a^2}\phi_0^\prime\phi^\prime_1+V_{1,\chi}\chi_1 \,.
\end{equation} 
Next, we define the density contrast as $\tilde\delta_\chi=\delta\tilde\rho_\chi/\tilde\rho_\chi$ and a new set of dimensionless variables 
\begin{align}
    &\sqrt{\frac{2}{3}}\frac{\kappa\,\chi^\prime_1}{\mathcal{H}}=-\tilde{\Omega}_\chi^{1/2}e^\alpha\cos\left(\frac{\vartheta}{2}\right), \\
    &\frac{\kappa\,y\,\chi_1}{\sqrt{6}}=-\tilde{\Omega}_\chi^{1/2}e^\alpha\sin\left(\frac{\vartheta}{2}\right), \\
    &\tilde{\delta}_\chi=-e^\alpha\sin\left(\frac{\theta-\vartheta}{2}\right), \\
    &\tilde{\Delta}_\chi=-e^\alpha\cos\left(\frac{\theta-\vartheta}{2}\right)\,,
\end{align}
where $e^\alpha$ represents the amplitude of perturbations, a placeholder that will ultimately drop out of the calculations. The quantity $\tilde\Delta_\chi$ is related to the pressure perturbation $\delta\tilde p_\chi$ so that
\begin{equation}
\delta\tilde p_\chi=\tilde\rho_\chi (\tilde\Delta_\chi\sin\theta-\tilde\delta_\chi\cos\theta).
\end{equation}
The new variables can also be related to the velocity divergence $\Theta$ given by Eq.~(\ref{divchi}) 
\begin{equation}
(\tilde\rho_\chi+\tilde p_\chi)\Theta=\frac{k^2\tilde\rho_\chi}{\mathcal{H}y}\Big[-\tilde\delta_\chi\sin\theta+\tilde\Delta_\chi(1-\cos\theta)\Big]\,.    
\end{equation}
So, one can see that by determining the values of the new variables $\tilde\delta_\chi$ and $\tilde\Delta_\chi$, we can fully compute the density and pressure perturbations as well as the velocity divergence of the field $\chi$. After a somewhat tedious calculation, we can write the evolution equations of these parameters as
\begin{align}
\tilde\delta^\prime_\chi&=\left[-\frac{3}{2}\left(1-\frac{\Omega_\chi}{\tilde\Omega_\chi}\right)(1-\cos\theta)+\frac{k^2}{\mathcal{H}^2y}\sin\theta\right]\mathcal{H}\tilde\delta_\chi-\frac{1}{2}h^\prime(1-\cos\theta) \nonumber \\
&+\left[-\frac{3}{2}\sin\theta\left(1+\frac{\Omega_\chi}{\tilde\Omega_\chi}\right)-\frac{k^2}{\mathcal{H}^2y}(1-\cos\theta)\right]\mathcal{H}\tilde\Delta_\chi+\frac{a^2}{\mathcal{H}y}\Big[\tilde\delta_\chi\sin\theta-\tilde\Delta_\chi(1-\cos\theta)\Big]V_{\text{int},\chi\chi} \nonumber \\
&-\sqrt{\frac{2}{3}}\frac{\kappa a^2}{\mathcal{H}\tilde\Omega_\chi^{1/2}}V_{\text{int},\chi\phi}\phi_1\sin\frac{\theta}{2}+\frac{\kappa a^2}{\sqrt{6}\mathcal{H}\tilde\Omega_\chi^{1/2}} V_{\text{int},\chi}\left(\tilde\delta_\chi \sin\frac{\theta}{2}-\tilde\Delta_\chi\cos\frac{\theta}{2}\right)\,,
\label{prt1}
\end{align}
and
\begin{align}
\tilde\Delta^\prime_\chi&=\left[-\frac{3}{2}\left(1-\frac{\Omega_\chi}{\tilde\Omega_\chi}\right)\sin\theta+\frac{k^2}{\mathcal{H}^2y}(1+\cos\theta)\right]\mathcal{H}\tilde\delta_\chi-\frac{1}{2}h^\prime\sin\theta \nonumber \\
&+\left[-\frac{3}{2}(1+\cos\theta)+\frac{3}{2}\frac{\Omega_\chi}{\tilde\Omega_\chi}(1-\cos\theta)-\frac{k^2}{\mathcal{H}^2y}\sin\theta\right]\mathcal{H}\tilde\Delta_\chi \nonumber \\
&+\frac{a^2}{\mathcal{H}y}\Big[\tilde\delta_\chi(1+\cos\theta)-\tilde\Delta_\chi\sin\theta\Big]V_{\text{int},\chi\chi}-\sqrt{\frac{2}{3}}\frac{\kappa a^2}{\mathcal{H}\tilde\Omega_\chi^{1/2}}V_{\text{int},\chi\phi}\phi_1\cos\frac{\theta}{2} \nonumber \\
&+\frac{\kappa a^2}{\sqrt{6}\mathcal{H}\tilde\Omega_\chi^{1/2}} V_{\text{int},\chi}\left(\tilde\delta_\chi\,\cos\frac{\theta}{2}+\tilde\Delta_\chi\sin\frac{\theta}{2}\right)\,.
\label{prt2}
\end{align}
We now have the complete set of DM and DE background and perturbation equations that can be solved along with the equations governing the evolution of the other species (baryons, photons and neutrinos).

\section{Constraining high scale parameters in SUGRA and Strings}

The analysis presented in this work also puts constraints on high scale parameters that appear in SUGRA and string models. Thus as noted in the beginning in SUGRA as well as in strings models one encounters an axion landscape. After breaking of the anomalous $U(1)_X$
global symmetry, the axion candidate for dark energy develops a potential which is a superposition of cosine terms of the form 
\begin{align}
V(a)= \mu^4 \sum_{k=1}^N  c_k \cos\left(\frac{k a}{F}\right), 
\end{align}
where $\mu$ carries the dimension of mass, and $c_k$ are dimensionless coefficients that depend on the details of the instanton breaking terms, and $F$ is the effective axion decay constant where 
\begin{align}
F= \sqrt{\sum_{k = 1}^{N} {f}_{k}^{2} + \sum_{k = 1}^{N} {\overline{f}}_{k}^{2} }\,.
\label{feF}
\end{align} 
Under the simplicity assumption that $f_k=\bar f_k=f$, one finds that 
$F= \sqrt{2 N} f$. In the analysis presented, we show that the cosmological data puts constraints on 
high scale string and SUGRA axionic model. Thus one finds that the data constrains the
effective axionic decay constant and also constrains the non-perturbative instanton 
corrections to the superpotential which break the anomalous $U(1)_X$ symmetry.
Thus as discussed in the Appendix the axionic potential for dark energy that arises from the instanton terms gives the potential
\begin{align}
{V} (a) &= \mu^4 \sum_{k=1}^N c_{k} \left(1+ \cos\left(\frac{k a}{F}\right)\right), \nonumber\\
 c_k &=  2 \sum_{r = 1}^{N} r  \left( {C}_{k  r} {f}_{k}^{r - 1}  \sum_{l = 1}^{N} l \, {C}_{k  l} {f}_{k}^{l - 1} + {\overline{C}}_{k  r}  {\overline{f}}_{k}^{r - 1} \sum_{l = 1}^{N} l\,  {\overline{C}}_{k  l} {\overline{f}}_{k}^{l - 1} \right),  
\label{slow2}
\end{align}
where, for simplicity, we have neglected the double sum term involving the coefficients $c_{rl}$. For illustration, we will consider the case $N=2$ and also use the simplifying assumption $\bar C_{kr}= C_{kr}$ and $\bar{f}_k^{r-1}= f_k^{r-1}$.  Under these assumptions we find 
\begin{align}
  \frac{c_2}{c_1}= \frac{C_{12}F  (C_{11} + C_{12} F) 
+  C_{22}F  (C_{21} + C_{22} F) }{C_{11}( C_{11} + C_{12} F) + C_{21} ( C_{21} +  C_{22} F)}\,.
 \end{align}

\section{Numerical analysis}

The background equations of DM, Eqs.~(\ref{bk1})$-$(\ref{bk3}), and DE, Eq.~(\ref{kgp0}), and the perturbation equations, Eq.~(\ref{prt1}), Eq.~(\ref{prt2}) and Eq.~(\ref{KGp1}) are solved, along with the evolution equations of the other species, using the Boltzmann solver \code{CLASS}~\cite{Blas:2011rf}. We consider three cases of Eq.~(\ref{v2}): the $N=1$ case
\begin{align}
    V_2(\phi)&=\mu^4\left[1+\cos\left(\frac{\phi}{F}\right)\right],
\end{align}
where $\mu^4c_1\to \mu^4$
and the $N=2$ case
\begin{align}
\label{v21}
V_2(\phi)&=\mu^4\left[1+\cos\left(\frac{\phi}{F}\right)\right]+\mu^4\left[1+\cos\left(\frac{2\phi}{F}\right)\right],
\end{align}
where we have set $c_2=c_1$ and
\begin{align}
\label{v22}
    V_2(\phi)&=\mu^4 c_1\left[1+\cos\left(\frac{\phi}{F}\right)\right]+\mu^4 c_2\left[1+\cos\left(\frac{2\phi}{F}\right)\right],
\end{align}
where $c_i$ are coefficients that are multiples of $\mu^4$. 
As noted in  Eq.~(\ref{v21}), we set $c_2=c_1$ and absorbed them in the definition of
$\mu^4$. In Eq.~(\ref{v22}), the coefficients $c_1$ and $c_2$ are different and we fix $\mu^4$. The $N=1$ case is the commonly used quintessence potential which, in this analysis, serves as a reference benchmark, while the $N=2$ case is the first non-trivial extension to this potential, inspired by supergravity and string models as discussed earlier. When it comes to the leading coefficients, $c_1$ and $c_2$, we also consider the most general case where the coefficients might be different and even opposite in sign. This can lead to constructive and destructive interference between cosine terms and potentially produce an interesting cosmology. We will test these notions in the MCMC analysis.

\subsection{Phenomenological study}

Before discussing the results of the MCMC simulations, we show the effect of the different forms of the potential $V_2(\phi)$ on the matter power spectrum and on the CMB power spectrum. 

\begin{figure}[t]
\begin{centering}
\includegraphics[width=0.49\linewidth]{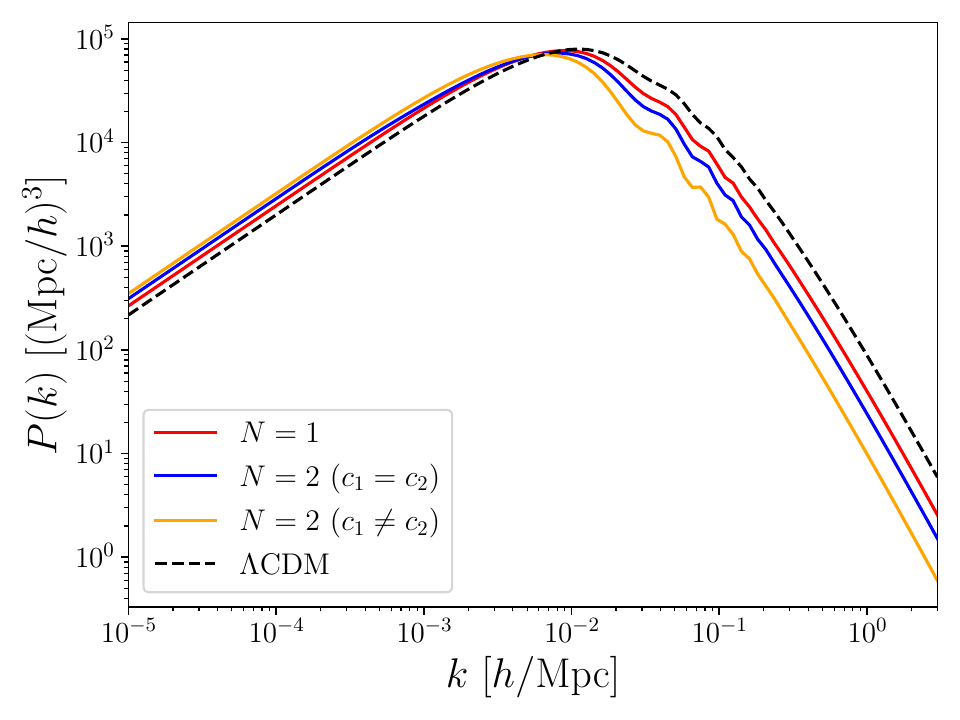}
\includegraphics[width=0.49\textwidth]{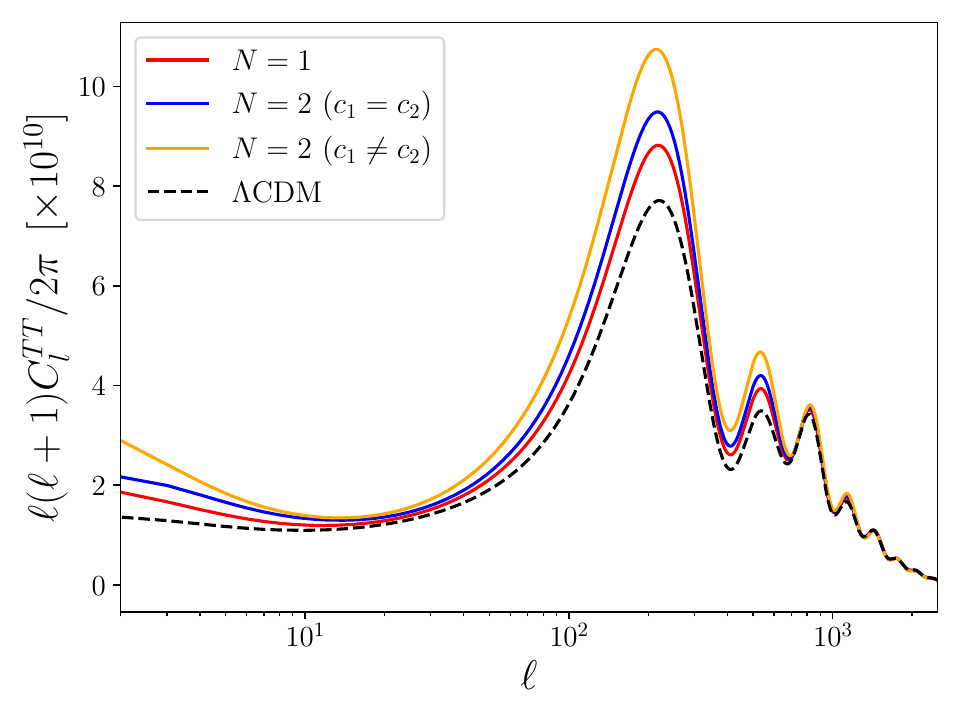}
\caption{The matter power spectrum versus the wavenumber $k$ (left panel) and the temperature TT power spectrum as a function of the multipoles (right panel) for three cases: $N=1$, $N=2$ with equal coefficients ($c_1=c_2=1$ with $\log\mu^4=-7.0$) and $N=2$ with different coefficients ($c_1=2.0$, $c_2=-0.1$ and $\log\mu^4=-7.0$). No DM-DE interaction is present. The $\Lambda$CDM case is shown as a dashed black curve.  }
\label{fig1}
\end{centering}
\end{figure}

The plots of Fig.~\ref{fig1} show the matter power spectrum (left panel) and the CMB temperature power spectrum (right panel) for three benchmarks: $N=1$, same-coefficient $N=2$ and different-coefficient $N=2$. The model shows suppression of the mass power spectrum at small scales in comparison to $\Lambda$CDM, while an enhancement is seen in the first and second acoustic peaks in the CMB temperature spectrum. The two effects can be attributed to a smaller matter density $\Omega_{\rm m}$ compared to $\Lambda$CDM. A smaller matter density means that matter-radiation equality happens later, thus allowing more modes entering the horizon during radiation domination to experience radiation driving. This driving leads to a simultaneous boost in the first and second acoustic peaks.

\begin{figure}[t]
\begin{centering}
\includegraphics[width=1.0\linewidth]{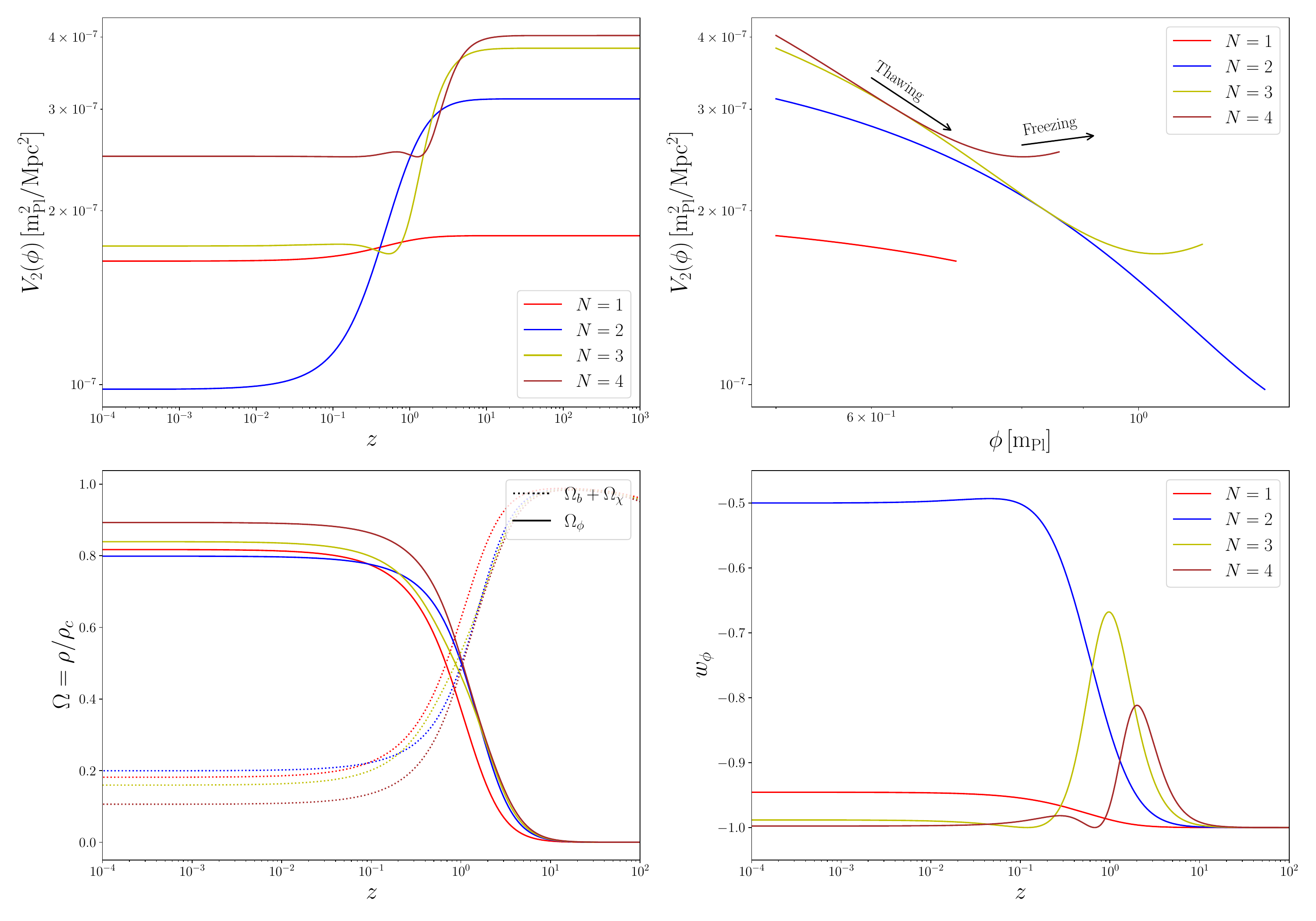}
\caption{Top panels: The dark energy potential $V_2(\phi)$ plotted against the redshift (left panel) and against the field $\phi$ (right panel) for the cases $N=1,2,3,4$ with equal coefficients, $c_i=1$ ($i=1,\cdots,4)$, and with $\log\mu^4=-7.0$. Bottom panels: the matter (dotted) and DE (solid) densities (left panel) and the DE equation of state (right panel) plotted against the redshift. No DM-DE interaction is present.}
\label{fig2}
\end{centering}
\end{figure}

To investigate the effect of superposition in the axion-like potential, we plot in the top panels of Fig.~\ref{fig2}, the variation of $V_2(\phi)$ with the redshift (left panel) and with the DE field $\phi$ (right panel) for $N=1,2,3,4$. For the $N=1$ case (red curve) we notice a standard thawing behavior of a quintessence field rolling slowly down the potential as well as the potential slowly dropping around $z\sim 1$ as the field gains kinetic energy. This causes the DE equation of state to depart away from $w_\phi=-1$ to higher values and one clearly see in the bottom right panel of Fig.~\ref{fig2}. The $N=2$ case brings about a more drastic behavior of the quintessence potential as shown in blue. The field rolls very rapidly over a small range of redshift causing a sudden and sharp departure of the EoS from $w_\phi=-1$, eventually leading to $w_\phi\simeq -0.5$ today. For higher $N$ values, $N=3,4$ (yellow and brown curves), the field begins rolling down the potential much rapidly than the $N=1$ case (top right panel), but then does something peculiar: it starts climbing up the potential, i.e., it transitions from a thawing to a freezing behavior. This is directly reflected in the EoS where a peak appears in the evolution of $w_\phi$ as it starts to dip again toward $w_\phi=-1$. The bottom left panel shows the evolution of the matter and DE density fractions with the redshift. There is an increasing DE density fraction at present $z$ starting from $N=3$ to $N=4$.

\subsection{MCMC results}

The Boltzmann solver \code{CLASS} is interfaced with \code{Cobaya}~\cite{Torrado:2020dgo}, a code for sampling and statistical modeling, to perform a Markov Chain Monte Carlo (MCMC) analysis of the model parameter space which allows us to extract constraints on the cosmological parameters through Bayesian inference. \code{Cobaya} utilizes an adaptive, speed-hierarchy-aware MCMC sampler (adapted from \code{CosmoMC})~\cite{Lewis:2002ah,Lewis:2013hha} using the fast-dragging procedure described in ref.~\cite{Neal:2005uqf}. We monitor the convergence of the chains using the Gelman-Rubin~\cite{Gelman:1992zz} criterion $R-1<0.05$. After convergence, the chains are analyzed with \code{GetDist}~\cite{Lewis:2019xzd}\footnote{\url{https://github.com/cmbant/getdist}}, a package allowing for the extraction of numerical results, including 1D posteriors and 2D marginalized probability contours.

In our analysis, we use the following data sets:
\begin{enumerate}
\item The temperature anisotropies and polarization measurements from the baseline high-$\ell$ TTTEEE \code{HiLLiPoP} likelihood from the Planck PR4 data release~\cite{Tristram:2020wbi,Tristram:2023haj} and the low-$\ell$ TT \code{Commander} likelihood~\cite{Planck:2018vyg}, supplemented with the low-$\ell$ EE \code{SimAll} likelihood~\cite{Planck:2018nkj,Planck:2019nip}.  In addition to the primary Planck CMB data, we also include measurements of the lensing potential power spectrum obtained from Planck \code{NPIPE} PR4 CMB maps~\cite{Carron:2022eyg}, along with lensing measurements from the Atacama Cosmology Telescope (ACT) made available with the Data Release 6 (DR6)~\cite{ACT:2023kun}. This whole data set is referred to as \textbf{CMB}.

\item Baryon Acoustic Oscillations (from DESI-DR2): BAO measurements from DESI's second data release (DR2), which takes into account observations of galaxies and quasars~\cite{DESI:2025zgx}, as well as Lyman-$\alpha$ tracers~\cite{DESI:2025zpo}, are included in the analysis. These measurements cover both isotropic and anisotropic BAO constraints over $0.295\leq z\leq 2.330$, divided into nine redshift bins. This data set is referred to as \textbf{DESI}.
\item The combination PantheonPlus+SH0ES~\cite{Brout:2022vxf,Riess:2021jrx} data set which uses an additional Cepheid distance as a calibrator of the Supernova SNIa intrinsic magnitude. This data set is referred to as \textbf{PPS}. 
\item Union 3.0: The Union 3.0 database, consisting of 2087 SN Ia within the range $0.001<z<2.260$~\cite{Rubin:2023jdq}. The data set has 1363 SN Ia in common with the PantheonPlus sample and uses a special and different treatment of systematic errors and uncertainties by employing Bayesian hierarchical modeling. We refer to this dataset as \textbf{Union3}.
\item DESY5: In their Year 5 data release, the Dark Energy Survey (DES) presented results based on a newly published, uniformly selected sample of 1635 photometrically classified Type Ia supernovae, covering redshifts from $0.1 < z < 1.3$~\cite{DES:2024jxu}. This sample is supplemented by 194 low-redshift SN Ia, drawn from the PantheonPlus dataset, within the range $0.025 < z < 0.1$. We refer to this combined dataset as \textbf{DESY5}.
\end{enumerate}

The sampling parameters consist of the baseline $\Lambda$CDM parameters and additional free parameters of our model. The $\Lambda$CDM parameters are
\begin{equation}
\Omega_{\rm b} h^2, ~~\Omega_\chi, ~~z_{\rm reio}, ~~100\theta_s, ~~\log(10^{10}A_s), ~~n_s\,,
\end{equation}
where $\Omega_\chi$ ($\Omega_{\rm b}$) is the DM (baryon) fractional density, $A_s$ is the amplitude of primordial fluctuations, $n_s$ is the spectral index of the primordial power spectrum, $\theta_s$ is the angular scale of the sound horizon at the surface of last scattering, and $z_{\rm reio}$ is the redshift at reionization. For the $N=1$ case, and the $N=2$ case with equal coefficients, the additional sampling parameters are
\begin{equation}
\log\mu^4,~~F,~~\phi_{\rm ini},~~\log\lambda\,,
\end{equation}
while for the $N=2$ case with different parameters, we have
\begin{align}
c_1,~~c_2,~~F,~~\phi_{\rm ini},~~\log\lambda\,.
\end{align}
We impose flat priors on all the parameters with ranges as shown in Table~\ref{tab-prior}.

\begin{table}[t]
\centering
{\tabulinesep=1.2mm
\resizebox{0.3\textwidth}{!}{\begin{tabu}{cc}
\hline\hline
\textbf{Parameter} & \textbf{Prior} \\
\hline
{$\log(10^{10} A_\mathrm{s})$} & $[1.61, 3.91]$ \\
{$n_\mathrm{s}$}  & $[0.8,1.2]$  \\
{$100\theta_\mathrm{s}$} & $[0.5,10]$ \\
{$\Omega_\mathrm{b} h^2$} & $[0.005,0.1]$ \\
{$\Omega_\chi$} & $[0.10,0.99]$ \\
{$\log\mu^4$}  & $[-9.0,-6.0]$ \\
{$F~[m_{\rm Pl}]$} & $[0.1,1.0]$ \\
{$\phi_{\mathrm{ini}}~[m_{\rm Pl}]$} & $[0.01,0.99]$ \\
{$\log\lambda$} & $[-8.0,-1.0]$ \\
$z_{\rm reio}$ & $[5.0, 9.5]$ \\
\hline\hline
\end{tabu}}}
\caption{The ranges of the flat priors adopted for the free cosmological parameters in the analysis of the QCDM model.}
\label{tab-prior}
\end{table}

\begin{table}[t]
\centering
{\tabulinesep=1.2mm
\resizebox{0.9\textwidth}{!}{\begin{tabu}{cccc}
\hline\hline
\textbf{Parameter} & \textbf{CMB+DESI+PPS} & \textbf{CMB+DESI+DESY5} & \textbf{CMB+DESI+Union3} \\
\hline
{$\log(10^{10} A_\mathrm{s})$} & $3.046\pm 0.013            $ & $3.044\pm 0.011            $ & $3.045\pm 0.012            $\\
{$n_\mathrm{s}   $}  & $0.9703\pm 0.0033          $ & $0.9696\pm 0.0037          $ & $0.9694\pm 0.0035          $\\
{$100\theta_\mathrm{s}$} & $1.04227\pm 0.00027        $ & $1.04219\pm 0.00028        $ & $1.04220\pm 0.00028        $\\
{$\Omega_\mathrm{b} h^2$} & $0.02283\pm 0.00011        $ & $0.02278\pm 0.00011        $ & $0.02279\pm 0.00011        $\\
{$\Omega_\chi    $} & $0.2415\pm 0.0066          $ & $0.256^{+0.011}_{-0.0097}  $ & $0.2543^{+0.0093}_{-0.012} $\\
{$\log\mu^4      $}  & $-7.230^{+0.013}_{-0.024}  $ & $-7.192^{+0.041}_{-0.065}  $ & $-7.175^{+0.053}_{-0.075}  $\\
{$F~[m_{\rm Pl}]$} & $> 0.620                   $ & $0.51^{+0.20}_{-0.22}      $ & $0.63^{+0.23}_{-0.17}      $\\
{$\phi_{\mathrm{ini}}~[m_{\rm Pl}]$} & $< 0.263                   $ & $< 0.384                   $ & $< 0.560                   $\\
{$\log\lambda    $} & $-5.49^{+0.97}_{-2.2}      $ & $< -5.49                   $ & $< -5.69                   $\\
\hline\hline
$H_0~[\rm km/s/Mpc]$ & $69.39\pm 0.87             $ & $67.6^{+1.3}_{-1.6}        $ & $67.8^{+1.6}_{-1.3}        $\\
$\Omega_\mathrm{m}         $ & $0.280^{+0.025}_{-0.011}   $ & $0.294^{+0.028}_{-0.015}   $ & $0.295^{+0.024}_{-0.015}   $\\
$\Omega_\phi               $ & $0.7097^{+0.0079}_{-0.0071}$ & $0.693\pm 0.012            $ & $0.695^{+0.014}_{-0.011}   $\\
$S_8                       $ & $0.773^{+0.028}_{-0.014}   $ & $0.778^{+0.027}_{-0.014}   $ & $0.780^{+0.023}_{-0.012}   $\\
$w_0                       $ & $-0.937^{+0.021}_{-0.17}   $ & $-0.925^{+0.055}_{-0.13}   $ & $-0.890^{+0.075}_{-0.16}   $\\
$w_a                       $ & $-0.018^{+0.029}_{-0.013}  $ & $-0.076^{+0.11}_{-0.046}   $ & $-0.109^{+0.14}_{-0.070}   $\\
\hline
$\Delta\mathrm{AIC}$ & 29.61 & 22.23 & 28.50 \\
\hline\hline
\end{tabu}}}
\caption{Constraints on some of the cosmological parameters of our model for the $N=1$ case. The values are quoted at 68\% CL intervals for three data set combinations. The middle double line separates the sampled and derived parameters using MCMC. In the last row we show the values of $\Delta\text{AIC}\equiv\text{AIC}_{\rm QCDM}-\text{AIC}_{\Lambda\rm CDM}$.  }
\label{tab1}
\end{table}

\begin{figure}[H]
\begin{centering}
\includegraphics[width=0.49\linewidth]{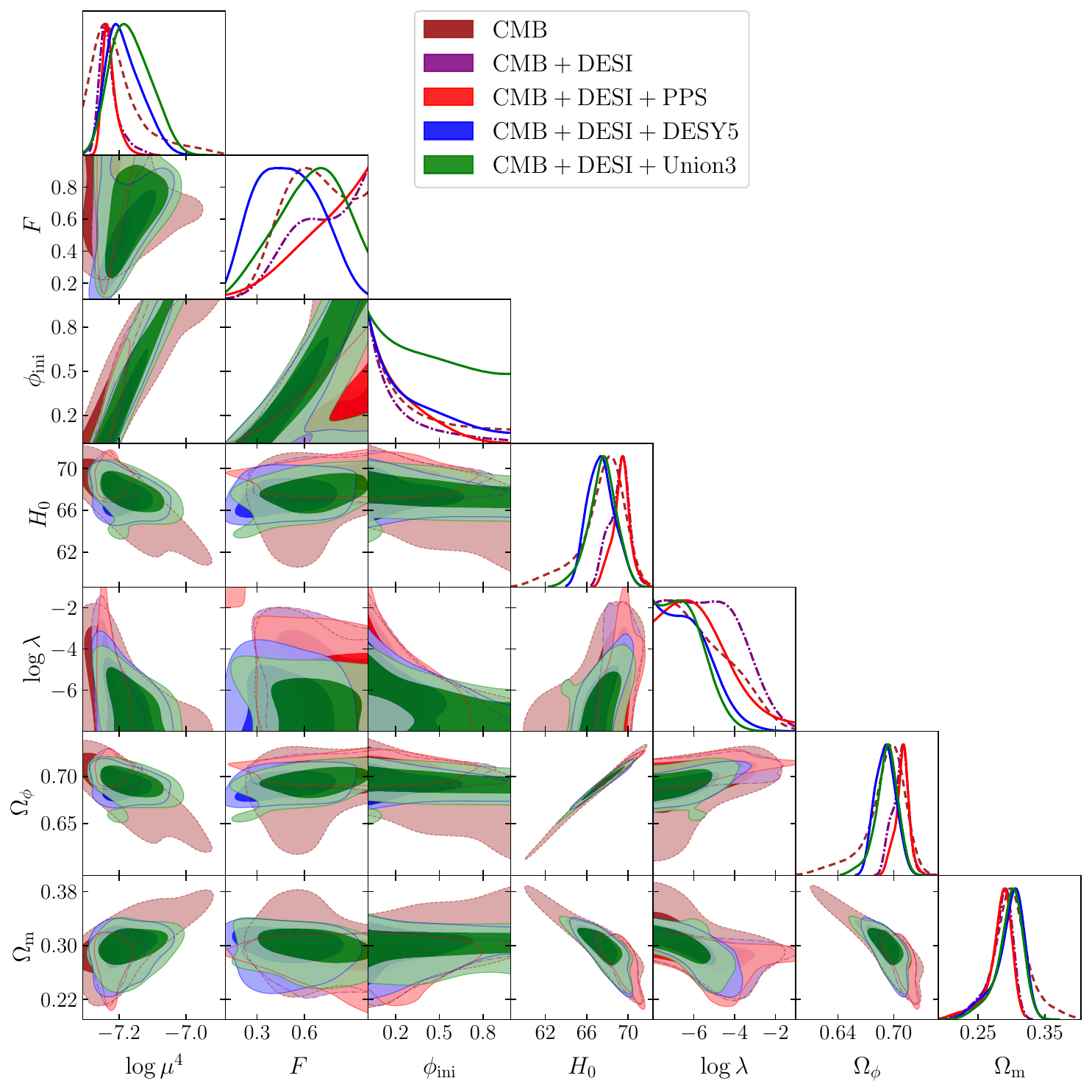}
\includegraphics[width=0.49\linewidth]{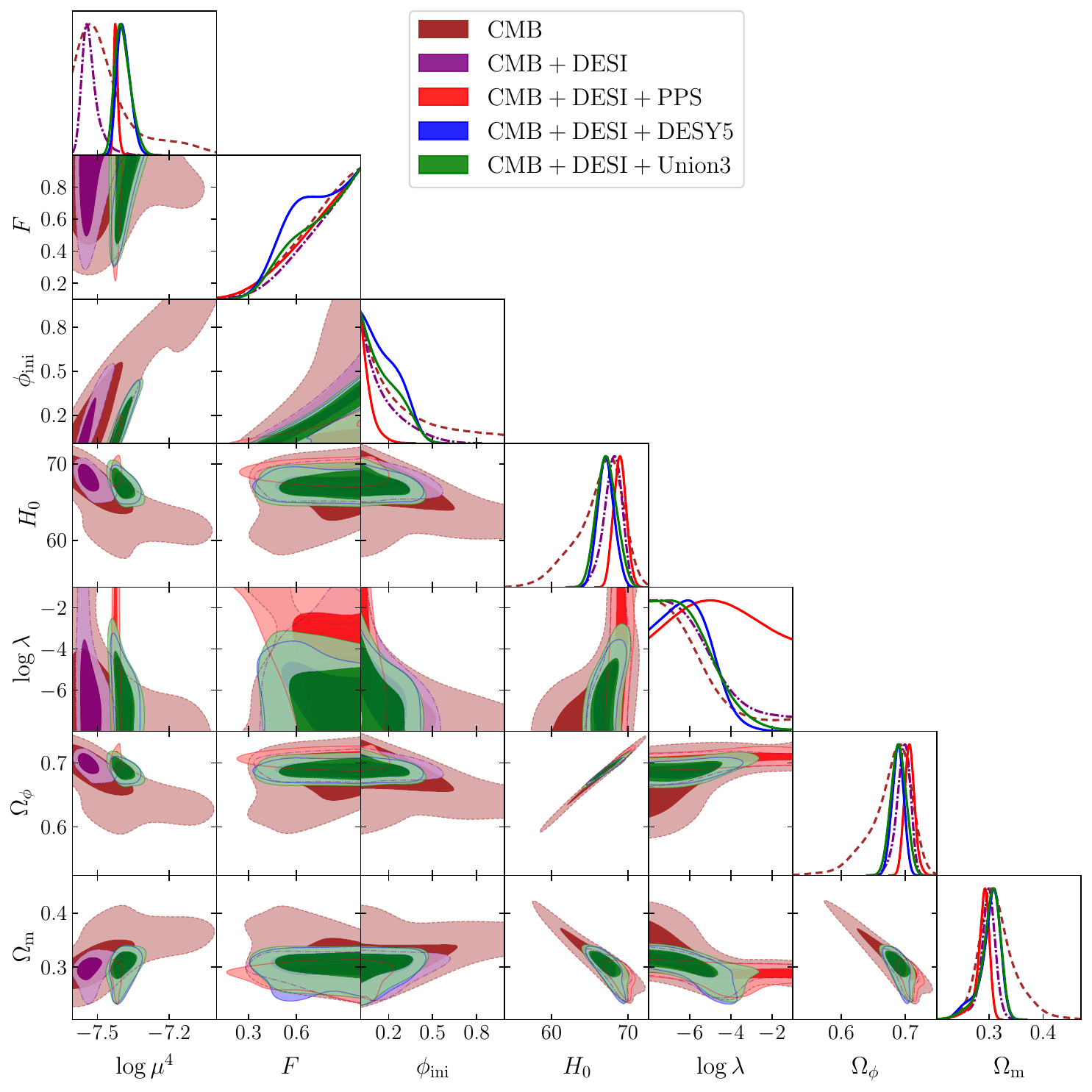} \\
\includegraphics[width=0.49\linewidth]{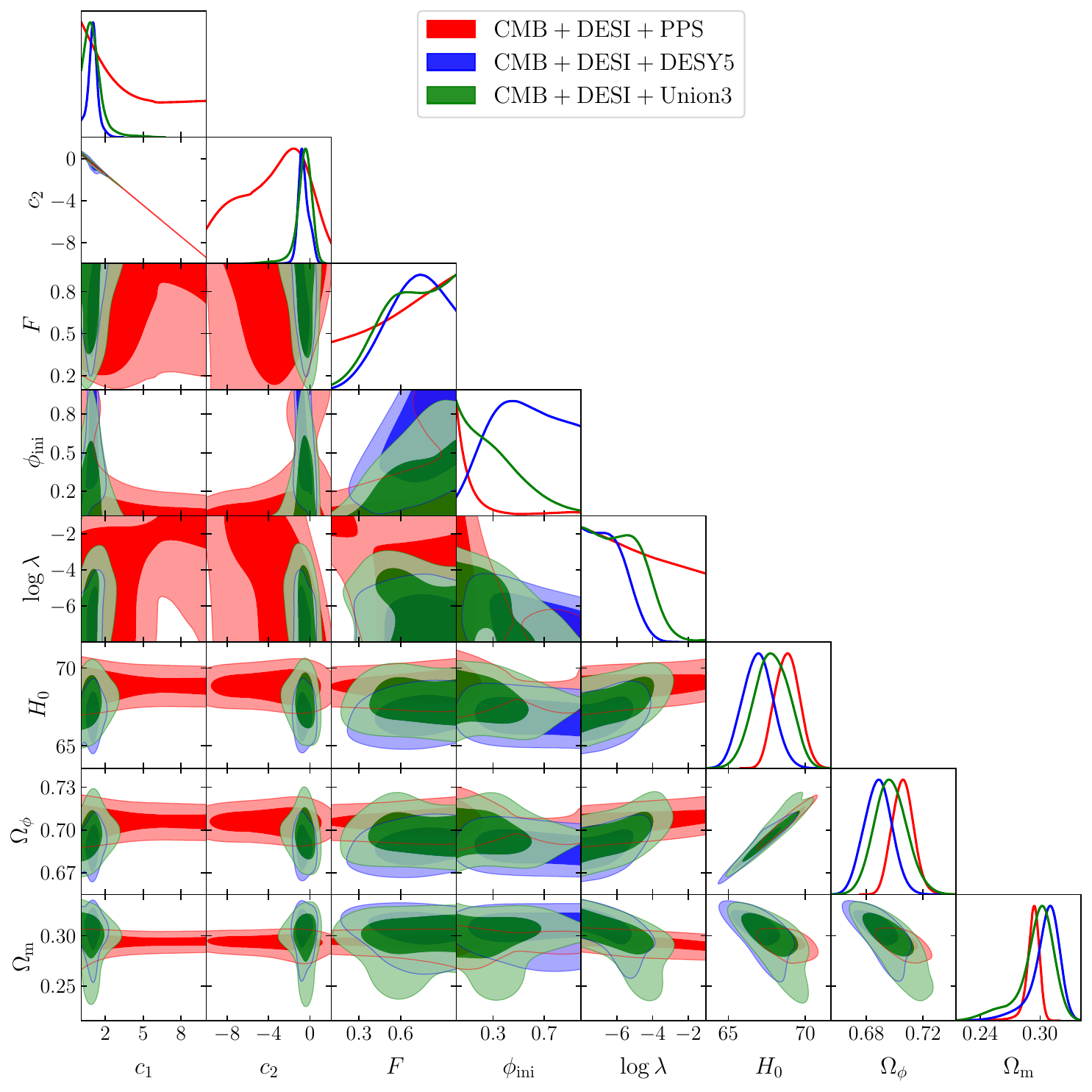}
\caption{The 1D and 2D marginalized posteriors for the $N=1$ (top left panel) and $N=2$ (top right panel) cases, showing correlations between different cosmological parameters for the five data sets. Here $c_1=c_2=1$. Bottom panel: same as top right panel but for $c_1\neq c_2$ and $\log\mu^4=-7.0$ while using only three data set combinations.  }
\label{fig3}
\end{centering}
\end{figure}

We begin the discussion of the results with the $N=1$ case, which are summarized in the upper left panel of Fig.~\ref{fig3}, as 1D and 2D marginalized posteriors of cosmological parameters, and in table~\ref{tab1}. The parameters in the upper part of the table (above the dividing double line) are the input parameters which we sample over and the parameters in the lower part are the derived ones. The baryon density $\Omega_{\rm b}h^2$ is almost the same for the three data sets, while the DM density $\Omega_\chi$ is smallest for the CMB+DESI+PPS data set and highest for CMB+DESI+DESY5. This can be understood based on the behavior of the derived parameters under the CMB+DESI+PPS data set that drives the Hubble parameter $H_0$ to a higher value which is reflected in a higher DE density $\Omega_\phi$ and thus a lower $\Omega_\chi$. We can see the strong anti-correlation between the matter density $\Omega_{\rm m}$ and $\Omega_\phi$ in Fig.~\ref{fig3}.  The model-dependent parameter $\mu^4$ takes values between $\sim 5.88\times 10^{-8}$ m$_{\rm Pl}^2$/Mpc$^2$ to $\sim 6.68\times 10^{-8}$ m$_{\rm Pl}^2$/Mpc$^2$ which is a very narrow range. The axion parameter $F$ is also constrained to within the range $\sim 0.30-0.86$ m$_{\rm Pl}$ for the DESY5 and Union3 data sets while only a lower limit is derived from the PPS data set. These $F$ values are sub-Planckian and are therefore consistent with conjectures from string theory that $F$ cannot be trans-Planckian. The analysis also sets an upper limit on $\phi_{\rm ini}$ and the most stringent of these constraints comes from the PPS data set. There are strong positive correlations between the parameters $\log\mu^4$, $F$ and $\phi_{\rm ini}$. Overall, the DM-DE interaction strength, $\lambda$, is weak with $\lambda\lesssim 3.2\times 10^{-6}$ m$_{\rm Pl}^{-2}$ Mpc$^{-2}$ with a mild correlation with $H_0$ for higher values. Regarding the Hubble tension, the $N=1$ case only provides a mild relief in the tension for the CMB+DESI+PPS data set, while the results of the other two data sets are consistent with $\Lambda$CDM. The \code{KiDS}-Legacy collaboration~\cite{Wright:2025xka} has released their final results of the $S_8$ parameter, the root-mean-square mass fluctuation amplitude for spheres of size $8h^{-1}$Mpc, at $S_8=0.815^{+0.016}_{-0.021}$. The values of $S_8$ as predicted by our model is consistent with the \code{KiDS}-Legacy data.

The dark energy equation of state is parametrized as
\begin{align}
    \hat{w}(z)&=-1+\alpha\,e^{-\beta/(1+z)}\arctan\frac{p}{(1+z)^q},
    \label{fit-weak}
\end{align}
where $\alpha$, $\beta$, $p$ and $q$ are the fitting parameters. Expanding Eq.~(\ref{fit-weak}) up to first order in $z/(1+z)$, we get $w_0$ and $w_a$ so that
\begin{align}
\label{w0-weak}
    w_0&=-1+\alpha\,e^{\beta}\,\arctan(p)\,, \\
    w_a&=\frac{\alpha\,e^{-\beta}}{1+p^2}\Big[-p\,q+\beta(1+p^2)\arctan(p)\Big]\,.
    \label{wa-weak}
\end{align}
The DE EoS, $w_\phi(z)$, obtained from our model is fitted to 
\begin{equation}
\hat{w}(z)=w_0+\frac{z}{1+z}w_a\,,    
\end{equation}
by minimizing the $\chi^2$
\begin{equation}
    \chi^2=\sum_i\left(\frac{w_i(z)-\hat{w}_i(z)}{\sigma_i}\right)^2\,,
\end{equation}
and extracting the values of $w_0$ and $w_a$ of each model data point. The 68\% CL values of $w_0$ and $w_a$ are summarized in table~\ref{tab1} for the $N=1$ case. The values are consistent with $\Lambda$CDM, with $w_0= -1$ and $w_a=0$, but their central values suggest a mild evolving DE equation of state. Unlike the DESI's interpretation~\cite{DESI:2025zgx} that shows that a $w_0w_a$CDM model exacerbates the Hubble tension for the data sets CMB+DESI+DESY5 and CMB+DESI+Union3, our model does not.

To find out whether our QCDM model fits the data better than $\Lambda$CDM, we use the Akaike Information Criterion (AIC)~\cite{Akaike:1974vps} which is defined as
\begin{equation}
    \text{AIC}\equiv -2\ln\mathcal{L}_{\rm max}+2K,
\end{equation}
where $\mathcal{L}_{\rm max}$ is the maximum likelihood of the model, and $K$ is the number of free parameters. The difference in AIC is defined as $\Delta\text{AIC}\equiv\text{AIC}_{\rm QCDM}-\text{AIC}_{\Lambda\rm CDM}$. The following rules are adopted: $\Delta\text{AIC}<-5$ indicates a strong preference for QCDM over $\Lambda$CDM, while $\Delta\text{AIC}>10$ indicates a decisive preference for $\Lambda$CDM. The last row in table~\ref{tab1} shows the values of $\Delta\text{AIC}$ for the three data sets and it is clear that our analysis indicates a preference for $\Lambda$CDM over QCDM.

\begin{table}[t]
\centering
{\tabulinesep=1.2mm
\resizebox{0.9\textwidth}{!}{\begin{tabu}{cccc}
\hline\hline
\textbf{Parameter} & \textbf{CMB+DESI+PPS} & \textbf{CMB+DESI+DESY5} & \textbf{CMB+DESI+Union3} \\
\hline
{$\log(10^{10} A_\mathrm{s})$} & $3.046\pm 0.013            $ & $3.044^{+0.010}_{-0.011}   $ & $3.046\pm 0.012            $\\
{$n_\mathrm{s}   $} & $0.9704\pm 0.0032          $ & $0.9691\pm 0.0036          $ & $0.9704\pm 0.0034          $\\
{$100\theta_\mathrm{s}$} & $1.04225\pm 0.00028        $ & $1.04220\pm 0.00027        $ & $1.04225\pm 0.00028        $\\
{$\Omega_\mathrm{b} h^2$} & $0.02284\pm 0.00011        $ & $0.02278\pm 0.00011        $ & $0.02280\pm 0.00011        $\\
{$\Omega_\chi    $} & $0.2427^{+0.0058}_{-0.0074}$ & $0.2541^{+0.0083}_{-0.011} $ & $0.260^{+0.010}_{-0.0090}  $\\
{$\log\mu^4      $} & $-7.5364^{+0.0097}_{-0.017}$ & $-7.508^{+0.032}_{-0.052}  $ & $-7.482^{+0.036}_{-0.051}  $\\
{$F~[m_{\rm Pl}]$} & $> 0.691                   $ & $0.62^{+0.18}_{-0.22}      $ & $> 0.599                   $\\
{$\phi_{\mathrm{ini}}~[m_{\rm Pl}]$} & $< 0.127                   $ & $< 0.234                   $ & $< 0.337                   $\\
{$\log\lambda    $} & $< -4.59                   $ & $-5.5^{+1.4}_{-1.7}        $ & $< -6.03                   $\\
\hline\hline 
$H_0~[\rm km/s/Mpc]$ & $69.2^{+1.0}_{-0.80}       $ & $68.0^{+1.7}_{-1.4}        $ & $67.0^{+1.2}_{-1.4}        $\\
$\Omega_\mathrm{m}         $ & $0.284^{+0.020}_{-0.0091}  $ & $0.284^{+0.037}_{-0.025}   $ & $0.304^{+0.019}_{-0.012}   $\\
$\Omega_\phi               $ & $0.7081^{+0.0089}_{-0.0068}$ & $0.695^{+0.013}_{-0.010}   $ & $0.688^{+0.011}_{-0.012}   $\\
$S_8                       $ & $0.777^{+0.024}_{-0.012}   $ & $0.768^{+0.039}_{-0.024}   $ & $0.785^{+0.019}_{-0.011}   $\\
$w_0                       $ & $-0.907^{-0.063}_{-0.092}  $ & $-0.961^{+0.042}_{-0.065}  $ & $-0.924^{+0.058}_{-0.12}   $\\
$w_a                       $ & $-0.0085^{+0.013}_{-0.0068}$ & $-0.035^{+0.051}_{-0.027}  $ & $-0.056^{+0.061}_{-0.033}  $\\
\hline
$\Delta\mathrm{AIC}$ & 8.83 & $-0.52$ & 6.45 \\
\hline\hline
\end{tabu}}}
\caption{Constraints on some of the cosmological parameters of our model for the $N=2$ case, with $c_1=c_2=1$. The values are quoted at 68\% CL intervals for three data set combinations. The middle double line separates the sampled and derived parameters using MCMC. In the last row we show the values of $\Delta\text{AIC}\equiv\text{AIC}_{\rm QCDM}-\text{AIC}_{\Lambda\rm CDM}$.  }
\label{tab2}
\end{table}

We now turn our attention to the $N=2$ case and assume here that the coefficients of the potential are equal to $\mu^4$; in other words, we are working with Eq.~(\ref{v21}). The results are summarized in table~\ref{tab2} and the posterior distributions are presented in the top right panel of Fig.~\ref{fig3}. The $N=2$ case produces a tighter constraints on $\phi_{\rm ini}$ in comparison to $N=1$, while requiring a smaller $\log\mu^4$ value. The axion decay constant $F$ is in the range $0.40-0.80$ m$_{\rm Pl}$ from DESY5 and a lower limit of $\sim 0.60\,\,\text{m}_{\rm Pl}$ comes from Union3 and $~\sim 0.7\,\,\text{m}_{\rm Pl}$ from PPS. The matter density fraction is almost the same as predicted in the $N=1$ case but lower than the $\Lambda$CDM prediction. Constraints on the DM-DE coupling strength $\lambda$ shows that if such an interaction exits it must be weak, but nonetheless can have some influence on $H_0$. In the top right panel of Fig.~\ref{fig3}, we can clearly see a slight positive correlation between $H_0$ and $\log\lambda$ for CMB+DESI+PPS data set, but this contribution is not sufficient to resolve the Hubble tension. The last row of table~\ref{tab2} shows $\Delta\text{AIC}$, and in comparison to the $N=1$ case, the $N=2$ case provides a better fit to the data. However, for the PPS and Union3 data sets, $\Lambda$CDM still does a better job while QCDM for $N=2$ fits the data almost as good as $\Lambda$CDM (or slightly better). Recall that here we are considering identical coefficients ($c_1=c_2$) in the quintessence potential and it is a remarkable result that a potential made of superposition of two cosine terms can fit the data better.

\begin{table}[t]
\centering
{\tabulinesep=1.2mm
\resizebox{0.9\textwidth}{!}{\begin{tabu}{cccc}
\hline\hline
\textbf{Parameter} & \textbf{CMB+DESI+PPS} & \textbf{CMB+DESI+DESY5} & \textbf{CMB+DESI+Union3} \\
\hline
{$\log(10^{10} A_\mathrm{s})$} & $3.055\pm 0.012            $ & $3.054\pm 0.011            $ & $3.051\pm 0.011            $\\
{$n_\mathrm{s}   $} & $0.9733\pm 0.0032          $ & $0.9729\pm 0.0031          $ & $0.9723\pm 0.0035          $\\
{$100\theta_\mathrm{s}$} & $1.04212\pm 0.00023        $ & $1.04210\pm 0.00022        $ & $1.04209\pm 0.00024        $\\
{$\Omega_\mathrm{b} h^2$} & $0.022564\pm 0.000098      $ & $0.022519\pm 0.000096      $ & $0.022516\pm 0.000096      $\\
{$\Omega_\chi    $} & $0.2452^{+0.0068}_{-0.0061}$ & $0.2600\pm 0.0082          $ & $0.252^{+0.010}_{-0.0091}  $\\
{$F              $} & $> 0.489                   $ & $0.68^{+0.27}_{-0.12}      $ & $> 0.557                   $\\
{$\phi_{\mathrm{ini}}$} & $< 0.0874                  $ & $0.54\pm 0.25              $ & $< 0.377                   $\\
{$c_1            $} & $< 5.57                    $ & $1.03\pm 0.38              $ & $1.09^{+0.25}_{-0.88}      $\\
{$c_2            $} & $-3.4^{+3.9}_{-2.8}        $ & $-0.61^{+0.39}_{-0.59}     $ & $-0.55^{+0.74}_{-0.44}     $\\
{$\log\lambda    $} & $< -3.65                   $ & $< -5.95                   $ & $< -4.97                   $\\
\hline\hline 
$H_0                       $ & $68.82\pm 0.77             $ & $66.90\pm 0.97             $ & $67.8\pm 1.1               $\\
$\Omega_\mathrm{m}         $ & $0.2925^{+0.0073}_{-0.0045}$ & $0.306^{+0.016}_{-0.0089}  $ & $0.297^{+0.020}_{-0.011}   $\\
$\Omega_\phi               $ & $0.7057^{+0.0071}_{-0.0080}$ & $0.6881\pm 0.0096          $ & $0.697\pm 0.011            $\\
$S_8                       $ & $0.795^{+0.010}_{-0.0081}  $ & $0.797^{+0.016}_{-0.0092}  $ & $0.793^{+0.020}_{-0.011}   $\\
$w_0                       $ & $-0.60^{+0.60}_{-0.40}     $ & $-0.956^{+0.029}_{-0.068}  $ & $-0.963^{+0.059}_{-0.086}  $\\
$w_a                       $ & $-1.8^{+1.8}_{+1.6}        $ & $-0.060^{+0.093}_{-0.044}  $ & $-0.031^{+0.076}_{-0.057}  $\\
\hline
$\Delta\mathrm{AIC}$ & 20.70 & 9.48 & 17.58 \\
\hline\hline
\end{tabu}}}
\caption{Constraints on some of the cosmological parameters of our model for the $N=2$ case with $c_1\neq c_2$ and $\log\mu^4=7.0$. The values are quoted at 68\% CL intervals for three data set combinations. The middle double line separates the sampled and derived parameters using MCMC. In the last row we show the values of $\Delta\text{AIC}\equiv\text{AIC}_{\rm QCDM}-\text{AIC}_{\Lambda\rm CDM}$.  }
\label{tab3}
\end{table}

\begin{figure}[t]
\begin{centering}
\includegraphics[width=0.5\linewidth]{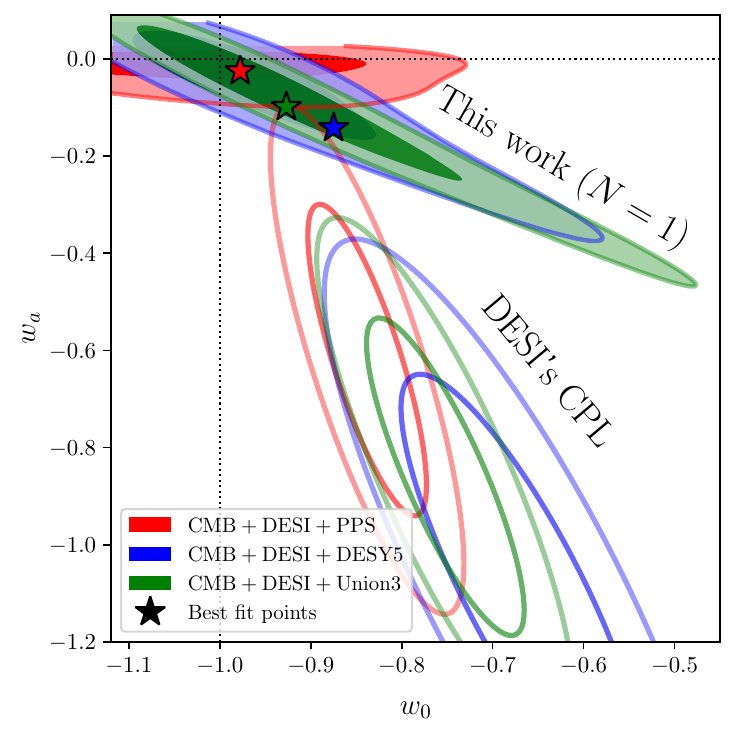}
\caption{Results for the posterior distributions of $w_0$ and $w_a$ for three data sets pertaining to the $N=1$ case. The best fit for each data set and DESI's contours are also shown. }
\label{fig4}
\end{centering}
\end{figure}

Next, we consider the $N=2$ case with different coefficients, $c_1\neq c_2$. Here we sample over these coefficients and fix $\log\mu^4=-7.0$. The results are shown in table~\ref{tab3} and the posteriors in the bottom panel of Fig.~\ref{fig3}. For the CMB+DESI+PPS data set, the DM-DE interaction strength has a higher limit which means that a stronger interaction can be accommodated by the data. This limit is the highest among all the cases considered. The positive correlation between $\log\lambda$ and $H_0$ is clear from the bottom panel of Fig.~\ref{fig3}, but again, this interaction does not result in a major relief in the Hubble tension. Furthermore, $\phi_{\rm ini}$ is constrained to very small values and the decay constant to values just above $\sim 0.5\,\,\text{m}_{\rm Pl}$. The quantity $w_a$ in the EoS is well constrained in this data set. For all the data sets, the analysis shows a baryon density higher than seen in the previous scenarios of tables~\ref{tab1} and~\ref{tab2} and a similar range for the decay constant $F$. In general, this case of $N=2$ with $c_1\neq c_2$ shows a higher $S_8$ value but still consistent with \code{KiDS}. The central values of the coefficient $c_2$ for the three data sets are negative while $c_1$ is positive. Having said that, it is clear that the data does not satisfactorily constrain these parameters. There is also a degradation in the value of $\Delta\text{AIC}$ compared to the same-coefficient $N=2$ case and this can be attributed to the additional free parameters that have been introduced.  

\begin{figure}[t]
\begin{centering}
\includegraphics[width=0.49\linewidth]{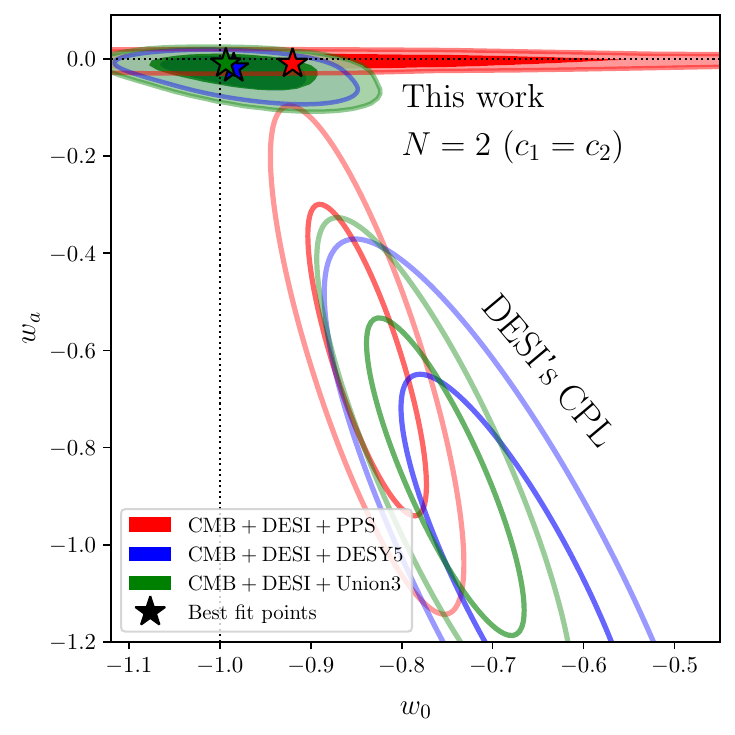}
\includegraphics[width=0.49\linewidth]{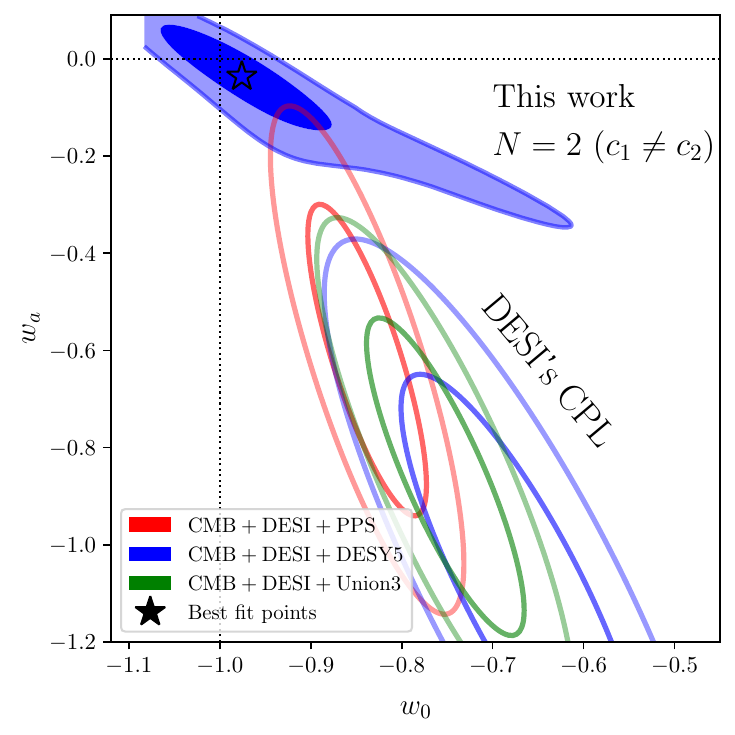}
\caption{Results for the posterior distributions of $w_0$ and $w_a$ for three data sets pertaining to the cases $N=2$ with equal coefficients ($c_1=c_2=1$) and $N=2$ with different coefficients ($c_1\neq c_2$). The best fit for each data set and DESI's contours are also shown. }
\label{fig5}
\end{centering}
\end{figure}

We now examine the idea of evolving dark energy EoS by determining the posterior distributions in the $w_0$-$w_a$ plane using Eqs.~(\ref{w0-weak}) and~(\ref{wa-weak}) for the three cases of $N$ considered in this analysis. The results are presented in Figs.~\ref{fig4} and~\ref{fig5} using the three data sets: CMB+DESI+PPS, CMB+DESI+DESY5 and CMB+DESI+Union3, along with the best fit points and the DESI's contours using the CPL parameterization. The $N=1$ case (Fig.~\ref{fig4}) shows the strongest scenario of evolving DE as the best fit points all lie in the fourth quadrant of the $w_0$-$w_a$ plane. The contours of our QCDM model lie above those of DESI's indicating a $w_a$ which is less negative and this is related to the fact that no crossing of the phantom divide happens in QCDM. The $N=2$ case with $c_1=c_2$ is shown in the left panel of Fig.~\ref{fig5}. Here, the best fit points are closer to $\Lambda$CDM and so this case does not provide an evidence of evolving dark energy EoS. The $N=2$ case with different coefficients is shown in the right panel of Fig.~\ref{fig5} where only one data set (CMB+DESI+DESY5) is exhibited and this is because the other two data sets give very poor constraints on $w_0$ and $w_a$.

The much broader error bars obtained with the data set CMB+DESI+PPS compared to CMB+DESI+DESY5/Union3, arise because Pantheon+SH0ES is in significant tension with the CMB- and BAO-calibrated distance scale, especially through the SH0ES $H_{0}$ prior, which favors $H_{0}\!\approx\!73~\mathrm{km\,s^{-1}\,Mpc^{-1}}$. In contrast, the CMB and DESI BAO measurements prefer $H_{0}\!\approx\!67$--$69~\mathrm{km\,s^{-1}\,Mpc^{-1}}$ and an expansion history close to $\Lambda$CDM. Since the CPL parametrization has only two degrees of freedom to absorb this multi-dataset tension, the joint fit is forced to wander along a large degeneracy direction in the $(w_{0},w_{a})$ plane, resulting in weak marginalized constraints such as the values of $w_0$ and $w_a$ shown in table~\ref{tab3} under CMB+DESI+PPS. Pantheon+SH0ES alone prefers a higher $H_{0}$, rapidly evolving dark-energy history, while CMB+DESI strongly disfavors this, and the combined likelihood therefore broadens instead of tightening. By contrast, the DESY5 and Union3 supernova compilations are much more consistent with CMB+DESI in both their calibrated distances and redshift-dependent systematics, leading to compact, nearly $\Lambda$CDM-like constraints. This behavior is consistent with recent analyses showing that Pantheon+SH0ES yields the weakest constraints on evolving dark energy when combined with BAO+CMB, while DESY5 and Union3 give the tightest and most internally consistent results. More concretely, refs.~\cite{Efstathiou:2024xcq,Giare:2024gpk} show that the DESY5 SN data drive a stronger apparent evolution in dark energy, whereas Pantheon+ SN data are consistent with no evolution. The preference for a $(w_0,w_a)\neq (-1,0)$ in combined fits is ``driven by the DESY5 SN", whereas ``the Pantheon+ catalogue is neutral towards evolving dark energy"~\cite{Efstathiou:2024xcq}. Therefore, it is not surprising that for the $N=2$ case with different coefficients, the only data set that is able to deliver an evolving DE is the DESY5 data set (right panel of Fig.~\ref{fig5}).

Finally, we note that these observational differences have prompted investigations into SN systematics. For instance, the DES collaboration finds that after accounting for calibration differences, the DESY5 evidence for evolving $w(z)$ remains $\sim 3.3\sigma$ (down from $3.9\sigma$)~\cite{DES:2025tir}, still higher than the Pantheon+ case.

\begin{table}[H]
\centering
\setlength{\tabcolsep}{3.5pt}
\renewcommand{\arraystretch}{1.05}
\scriptsize
\resizebox{\textwidth}{!}{%
\begin{tabular}{lcccccc}
\toprule
& \multicolumn{3}{c}{$N=1$} & \multicolumn{3}{c}{$N=2$} \\
\cmidrule(lr){2-4} \cmidrule(lr){5-7}
\textbf{Parameter} & \textbf{CMB} & \textbf{CMB+DESI} & \textbf{DESI+PPS} & \textbf{CMB} & \textbf{CMB+DESI} & \textbf{DESI+PPS} \\
\midrule
$\log(10^{10} A_\mathrm{s})$ 
& $3.048\pm 0.011$ & $3.042\pm 0.012$ & $< 2.75$
& $3.049\pm 0.013$ & $3.052\pm 0.012$ & $< 2.72$ \\

$n_\mathrm{s}$ 
& $0.9701\pm 0.0039$ & $0.9685\pm 0.0035$ & $< 0.952$
& $0.9699\pm 0.0040$ & $0.9721\pm 0.0032$ & $< 0.953$ \\

$100\theta_\mathrm{s}$ 
& $1.04200\pm 0.00024$ & $1.04217\pm 0.00028$ & $1.044^{+0.025}_{-0.015}$
& $1.04201\pm 0.00023$ & $1.04207\pm 0.00023$ & $1.042^{+0.027}_{-0.016}$ \\

$\Omega_\mathrm{b} h^2$ 
& $0.022497\pm 0.000095$ & $0.02278\pm 0.00011$ & $0.0260^{+0.010}_{-0.0083}$
& $0.022494\pm 0.000098$ & $0.022515\pm 0.000097$ & $0.0257^{+0.011}_{-0.0087}$ \\

$\Omega_\chi$ 
& $0.258^{+0.011}_{-0.021}$ & $0.2442^{+0.0064}_{-0.0082}$ & $0.249^{+0.012}_{-0.014}$
& $0.269^{+0.017}_{-0.031}$ & $0.2529^{+0.0076}_{-0.010}$ & $0.250^{+0.012}_{-0.015}$ \\

$\log\mu^4$ 
& $-7.207^{+0.036}_{-0.091}$ & $-7.230^{+0.016}_{-0.034}$ & $-7.15^{+0.11}_{-0.094}$
& $-7.473^{+0.048}_{-0.15}$ & $-7.531^{+0.021}_{-0.039}$ & $-7.47^{+0.11}_{-0.091}$ \\

$F~[m_{\rm Pl}]$ 
& $> 0.552$ & $> 0.588$ & $0.70^{+0.25}_{-0.12}$
& $> 0.682$ & $> 0.704$ & $> 0.725$ \\

$\phi_{\mathrm{ini}}~[m_{\rm Pl}]$ 
& $< 0.386$ & $< 0.246$ & ---
& $< 0.325$ & $< 0.200$ & $< 0.347$ \\

$\log\lambda$ 
& $< -5.08$ & $-5.35^{+0.86}_{-2.6}$ & $< -6.20$
& $< -5.76$ & $< -5.39$ & $< -6.01$ \\

\midrule

$H_0~[\mathrm{km/s/Mpc}]$ 
& $67.5^{+2.5}_{-1.3}$ & $69.1^{+1.1}_{-0.86}$ & $71^{+8}_{-6}$
& $66.3^{+3.5}_{-2.0}$ & $68.0^{+1.3}_{-1.0}$ & $70^{+8}_{-6}$ \\

$\Omega_\mathrm{m}$ 
& $0.299\pm 0.029$ & $0.283^{+0.022}_{-0.011}$ & $0.294^{+0.017}_{-0.010}$
& $0.315^{+0.025}_{-0.037}$ & $0.296^{+0.016}_{-0.011}$ & $0.296^{+0.016}_{-0.0098}$ \\

$\Omega_\phi$ 
& $0.691^{+0.024}_{-0.013}$ & $0.7067^{+0.0098}_{-0.0076}$ & $0.699\pm 0.010$
& $0.678^{+0.037}_{-0.020}$ & $0.697^{+0.012}_{-0.0090}$ & $0.698\pm 0.010$ \\

$S_8$ 
& $0.794^{+0.025}_{-0.015}$ & --- & $0.62^{+0.14}_{-0.24}$
& $0.803^{+0.020}_{-0.015}$ & $0.794^{+0.018}_{-0.010}$ & $0.61^{+0.14}_{-0.24}$ \\

$w_0$ 
& $-0.873^{+0.083}_{-0.26}$ & $-0.909^{+0.072}_{-0.21}$ & $-0.858^{+0.086}_{-0.18}$
& $-0.468^{+0.047}_{-0.53}$ & $-0.884^{+0.078}_{-0.24}$ & $-0.903^{+0.074}_{-0.14}$ \\

$w_a$ 
& $-0.091^{+0.18}_{-0.074}$ & $-0.024^{+0.042}_{-0.021}$ & $-0.107^{+0.11}_{-0.068}$
& $-0.44^{+0.43}_{+0.20}$ & $-0.033^{+0.051}_{-0.027}$ & $-0.067^{+0.070}_{-0.042}$ \\

\midrule
$\Delta\mathrm{AIC}$ 
& $20.50$ & $34.06$ & $12.28$
& 20.77 & 34.42 & 12.30 \\
\bottomrule
\end{tabular}%
}
\caption{Constraints on some of the cosmological parameters of our model for the $N=1$ and $N=2$ (with $c_1=c_2=1)$ cases from CMB only, CMB+DESI and the purely late-time data set, DESI+PPS. In the last row we show the values of $\Delta\mathrm{AIC}\equiv \mathrm{AIC}_{\rm QCDM}-\mathrm{AIC}_{\Lambda\rm CDM}$.}
\label{tab4}
\end{table}

\subsection{Discussion of tensions between data sets}

To assess the mutual consistency of the datasets within the extended model, we performed separate MCMC analyses for individual and partial combinations of the data in the $N=1$ case, namely CMB alone, CMB+DESI, and DESI+PPS (table~\ref{tab4}), before considering the full joint constraints (tables~\ref{tab1} and~\ref{tab2}). We find that the inferred posteriors for key cosmological parameters are statistically compatible across these datasets. In particular, the Hubble parameter shifts from $H_0 = 67.5^{+2.5}_{-1.3}\,\mathrm{km/s/Mpc}$ for CMB alone to $H_0 = 69.1^{+1.1}_{-0.86}\,\mathrm{km/s/Mpc}$ for CMB+DESI and $H_0 = 71^{+8}_{-6}\,\mathrm{km/s/Mpc}$ for DESI+PPS, with the corresponding intervals overlapping within $\sim 1$--$2\sigma$. This indicates that, although the late-time data prefer higher values of $H_0$, the datasets are not in strong tension within the model. A similar conclusion holds for the matter density, where $\Omega_{\mathrm{m}} = 0.299\pm 0.029$ (CMB), $0.283^{+0.022}_{-0.011}$ (CMB+DESI), and $0.294^{+0.017}_{-0.010}$ (DESI+PPS) are in good agreement. The dark energy parameters also exhibit consistent behavior, with $w_0$ ranging from $-0.873^{+0.083}_{-0.26}$ (CMB) to $-0.909^{+0.072}_{-0.21}$ (CMB+DESI) and $-0.858^{+0.086}_{-0.18}$ (DESI+PPS), and $w_a$ remaining close to zero within uncertainties in all cases. Upon combining the datasets, the constraints primarily tighten without inducing significant shifts in the central values. For example, $H_0 = 69.39\pm 0.87\,\mathrm{km/s/Mpc}$ for CMB+DESI+PPS compared to $67.6^{+1.3}_{-1.6}\,\mathrm{km/s/Mpc}$ and $67.8^{+1.6}_{-1.3}\,\mathrm{km/s/Mpc}$ for CMB+DESI +DESY5 and $\mathrm{CMB+DESI+Union3}$, respectively, indicating only a mild upward shift driven by PPS. From a physical perspective, this behavior reflects the limited freedom of the $N=1$ model to deviate from $\Lambda$CDM, with $\log\lambda$ constrained to be small (e.g. $\log\lambda = -5.49^{+0.97}_{-2.2}$) and the scalar field parameters restricted (e.g. $F > 0.62$), resulting in only mild late-time dynamics and a correspondingly modest alleviation of the Hubble tension.

A similar consistency analysis was carried out for the $N=2$ case, yielding qualitatively the same conclusions, with some additional features. The individual datasets again show statistically compatible constraints prior to combination. In particular, the Hubble parameter evolves from $H_0 = 66.3^{+3.5}_{-2.0}\,\mathrm{km/s/Mpc}$ (CMB) to $68.0^{+1.3}_{-1.0}\,\mathrm{km/s/Mpc}$ (CMB+DESI) and $70^{+8}_{-6}\,\mathrm{km/s/Mpc}$ (DESI+PPS), with overlapping uncertainties indicating no significant residual tension. The matter density remains consistent across datasets, with $\Omega_{\mathrm{m}} = 0.315^{+0.025}_{-0.037}$ (CMB), $0.296^{+0.016}_{-0.011}$ (CMB+DESI), and $0.296^{+0.016}_{-0.0098}$ (DESI+PPS). The dark energy parameters further illustrate this consistency: while CMB alone provides weak constraints ($w_0 = -0.468^{+0.047}_{-0.53}$, $w_a = -0.44^{+0.43}_{+0.20}$), the addition of DESI significantly tightens the parameter space to $w_0 \simeq -0.88$ and $w_a \simeq 0$, consistent with DESI+PPS results ($w_0 = -0.903^{+0.074}_{-0.14}$, $w_a = -0.067^{+0.070}_{-0.042}$). The full joint constraints again show only mild shifts, with $H_0 = 69.2^{+1.0}_{-0.80}\,\mathrm{km/s/Mpc}$ for CMB+DESI+PPS compared to $68.0^{+1.7}_{-1.4}\,\mathrm{km/s/Mpc}$ and $67.0^{+1.2}_{-1.4}\,\mathrm{km/s/Mpc}$ for the DESY5 and Union3 combinations, respectively, all consistent within $\sim 1$--$2\sigma$. Notably, the $N=2$ model exhibits slightly improved statistical performance in some cases, e.g. $\Delta\mathrm{AIC} = -0.52$ for CMB+DESI+DESY5, indicating that the additional scalar degree of freedom can provide a marginally better fit without introducing inconsistencies. Physically, this reflects the increased flexibility of the scalar sector, allowing for mild late-time evolution that accommodates both early- and late-Universe constraints while maintaining overall consistency. Nevertheless, the constraints remain close to $\Lambda$CDM, with parameters such as $\log\lambda$ still restricted to small values (e.g. $\log\lambda < -4.59$ or tighter), indicating that any deviations from standard cosmology are modest and do not generate significant tensions between the datasets.

\section{Conclusion}

In this work, a model of dark energy arising from a quintessence axionic field with a potential which is a superposition of $N$ number of cosine terms motivated by high scale string inspired SUGRA models is discussed. Numerical analyses for the cases $N=1,2$ are presented in detail using a Markov Chain Monte Carlo simulation with the Planck+DESI+PantheonSH0ES /DESY5/Union3 data sets. The analysis shows that the $N=2$ case constitutes a better fit to the data than the standard $N=1$ case, and their relative deviations from the $\Lambda$CDM model are also discussed. The analysis puts strong constraints on the high scale parameters and specifically on the axion decay constant $F$ which  is determined to be sub-Planckian but close to the Planck scale. The DM-DE interaction strength $\lambda$ is also constrained with an upper limit of $\lambda\lesssim 4\times 10^{-6}$ m$_{\rm Pl}^{-2}$ Mpc$^{-2}$. The $N=3,4$ cases bring in a scenario the authors have discussed in an earlier publication~\cite{Aboubrahim:2024cyk} which is the transmutation of quintessence between thawing and freezing. However, in the previous work, this transmutation occurred as a result of an appreciable DM-DE interaction. Here, the transmutation happens solely due to the superposition of axionic terms in the potential. Thus, the field starts rolling down its potential and the EoS begins to deviate away from $-1$ before the field starts climbing back up the potential as it loses kinetic energy. As a consequence, the EoS plunges back toward $-1$.
Finally we note that in many previous analyses which work at the level of continuity equations for dark energy and dark matter, the sources are arbitrarily chosen to satisfy energy conservation. However, it is shown in~\cite{Aboubrahim:2024cyk} that such analyses are inconsistent unless the source term is zero, i.e., no interaction between DE and DM. In contrast, the analysis presented here within QCDM is internally consistent with the sources for dark energy and dark matter in continuity equations being automatically determined by the underlying Lagrangian.

\begin{figure}[t]
\begin{centering}
\includegraphics[width=0.49\linewidth]{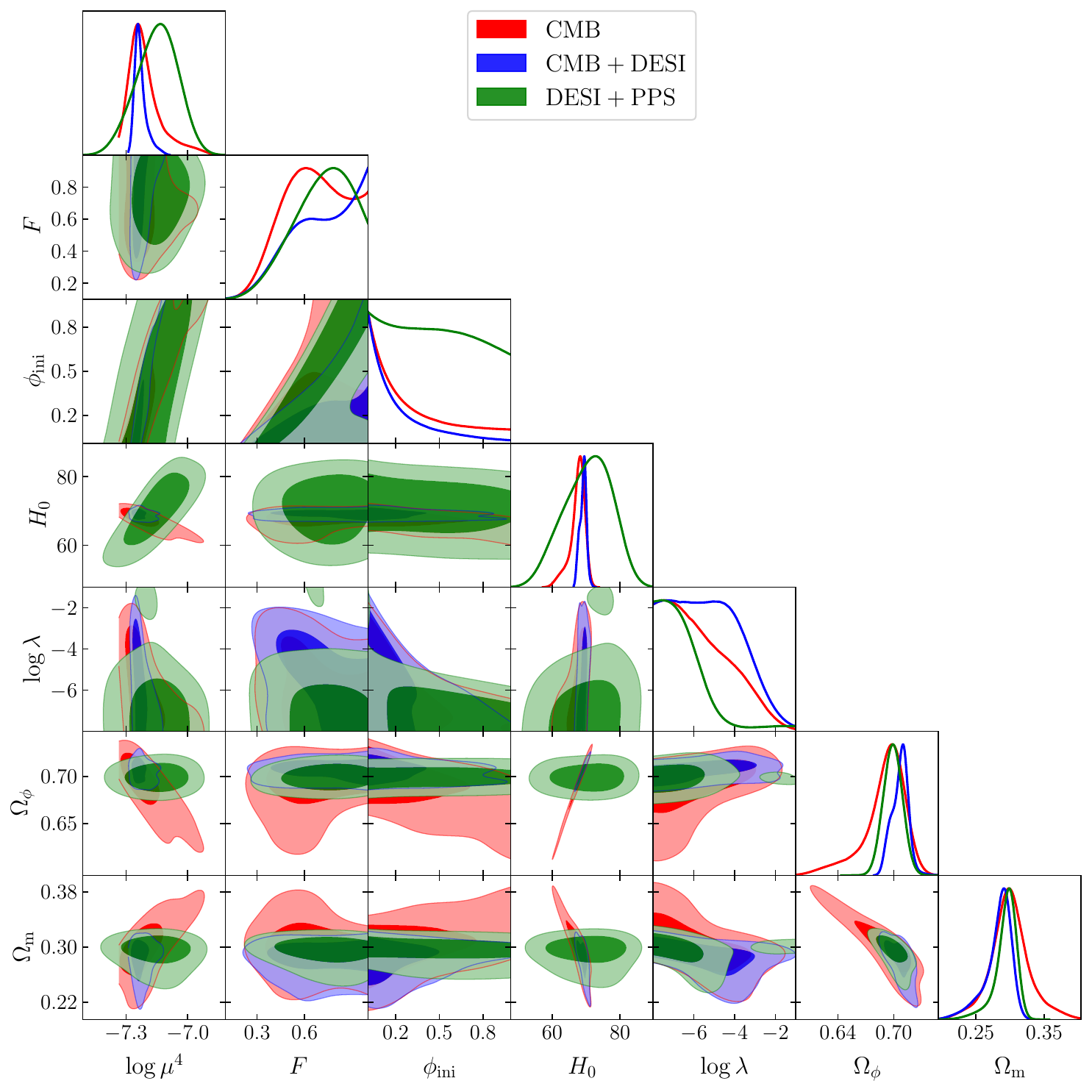}
\includegraphics[width=0.49\linewidth]{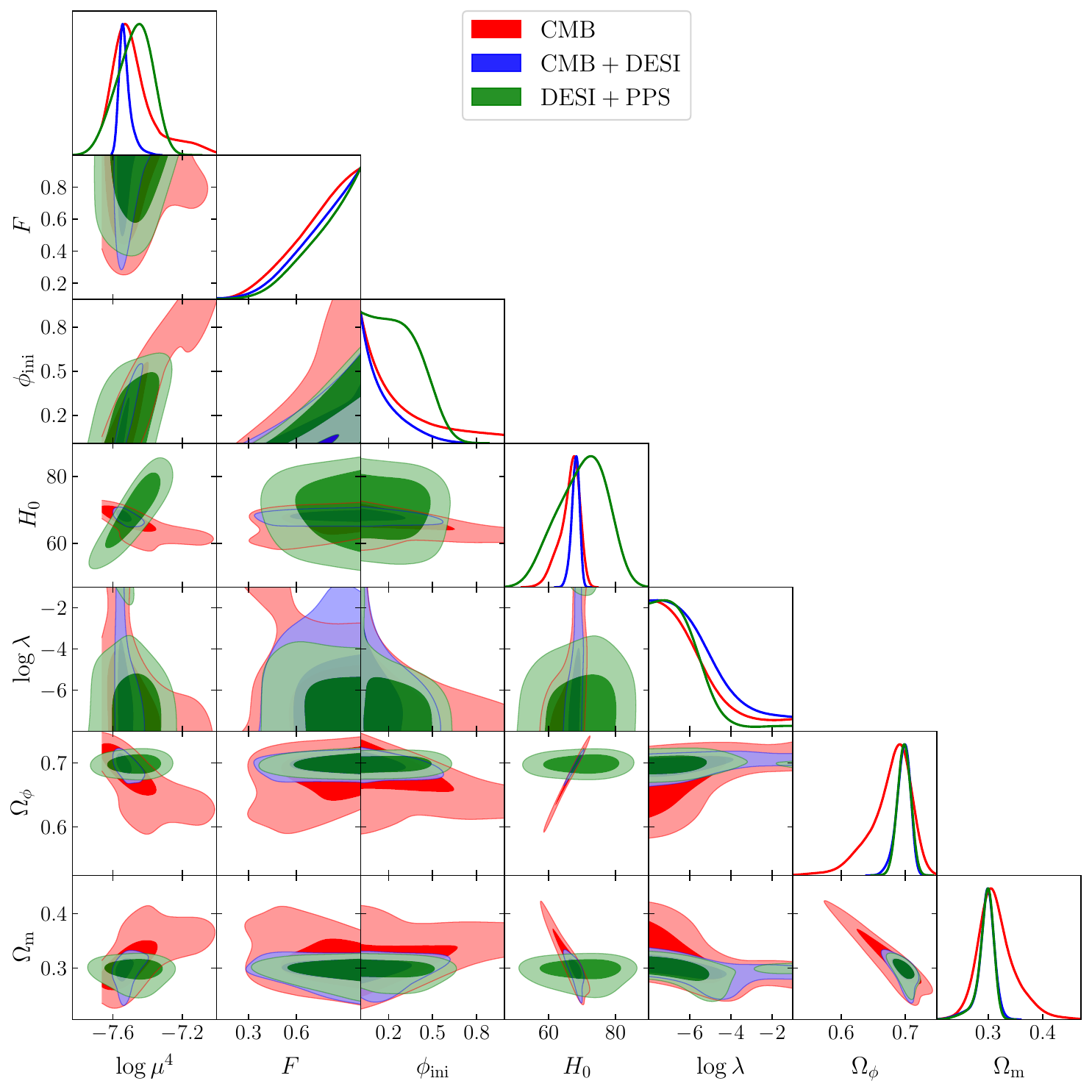} 
\caption{The 1D and 2D marginalized posteriors for the $N=1$ (left panel) and $N=2$ (right panel) cases, showing correlations between different cosmological parameters for three data sets: CMB only, CMB+DESI and the purely late-time data set, DESI+PPS. For the $N=2$ case, we have $c_1=c_2=1$. }
\label{fig6}
\end{centering}
\end{figure}

\vspace{1cm}

{\bf Acknowledgments:} 
The research of PN was supported in part by the NSF Grant PHY-2209903. This work used the Anvil Cluster at Purdue University through allocation PHY250220 from the \textbf{Advanced Cyberinfrastructure Coordination Ecosystem: Services \& Support} (ACCESS) program, which is supported by U.S. National Science Foundation grants \#2138259, \#2138286, \#2138307, \#2137603, and \#2138296.

\appendix

\section*{Appendix}

We present in this appendix the full 1D and 2D posterior distributions for the $N=1$ and $N=2$ cases considered in the analysis.

\begin{figure}[H]
\begin{centering}
\includegraphics[width=1.0\linewidth]{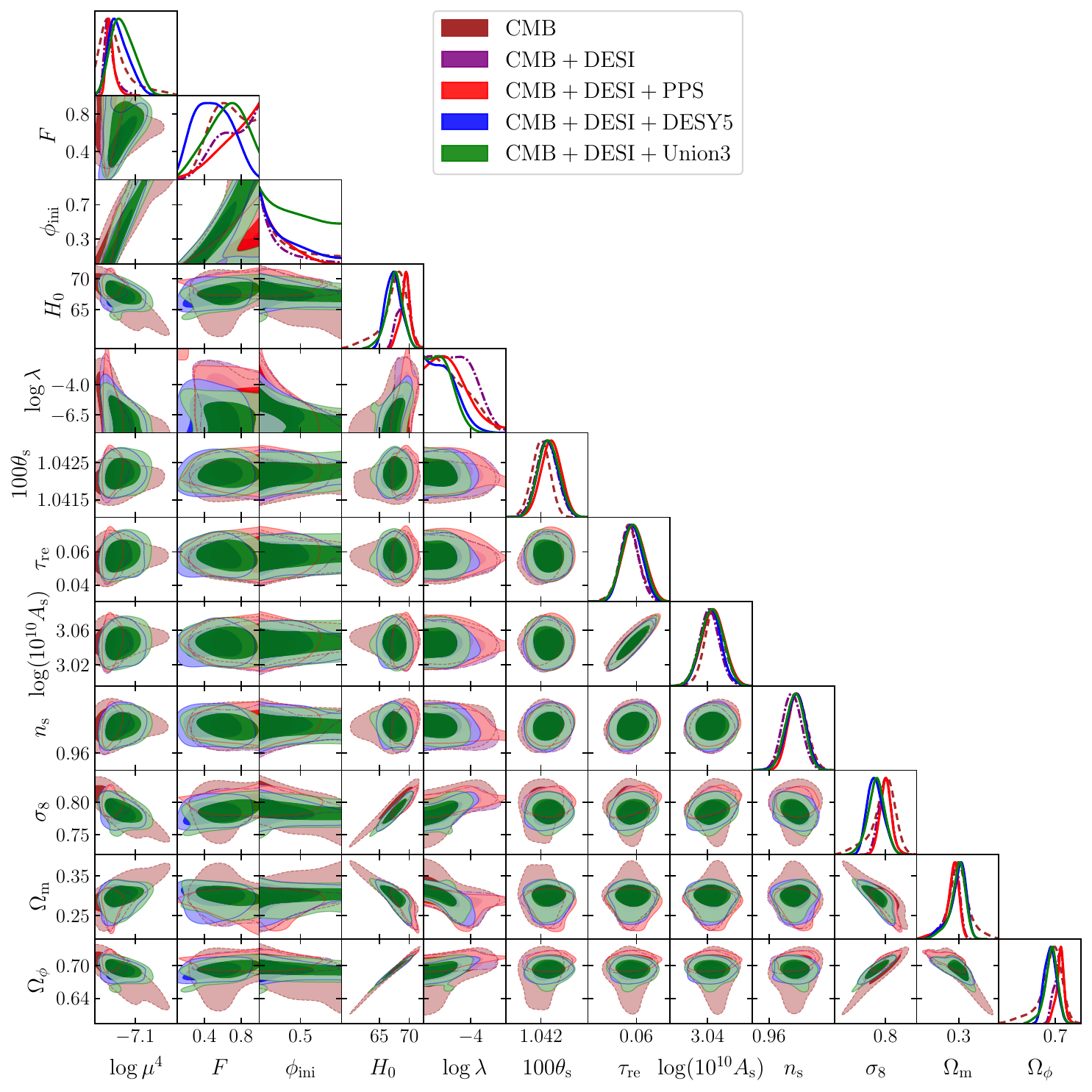}
\caption{The 1D and 2D marginalized posteriors for the $N=1$ case.  }
\label{fig1A}
\end{centering}
\end{figure}

\begin{figure}[H]
\begin{centering}
\includegraphics[width=1.0\linewidth]{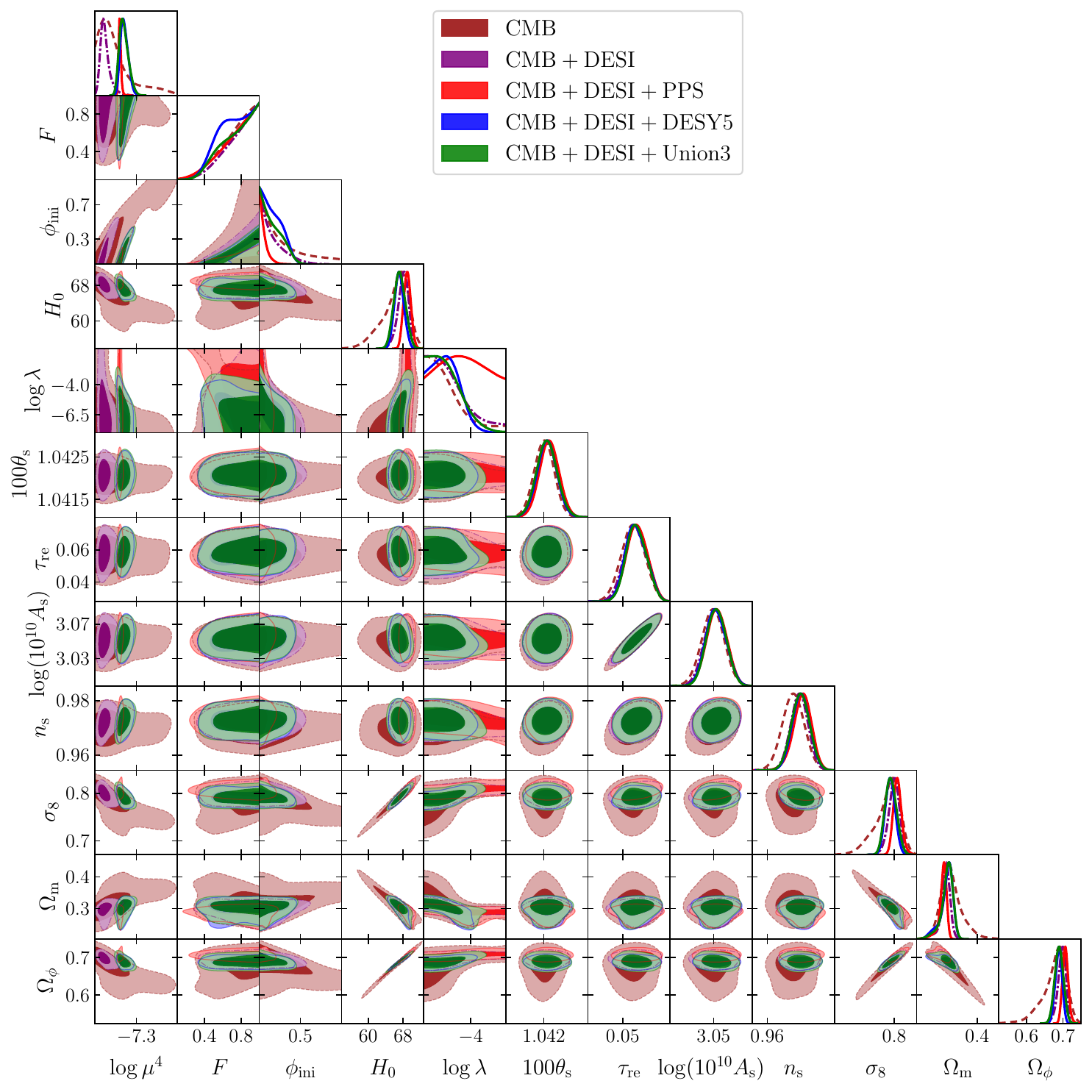}
\caption{The 1D and 2D marginalized posteriors for the $N=2$ case 
when $c_2=c_1$ and they are absorbed in $\mu^4$.
 }
\label{fig2A}
\end{centering}
\end{figure}

\begin{figure}[H]
\begin{centering}
\includegraphics[width=1.0\linewidth]{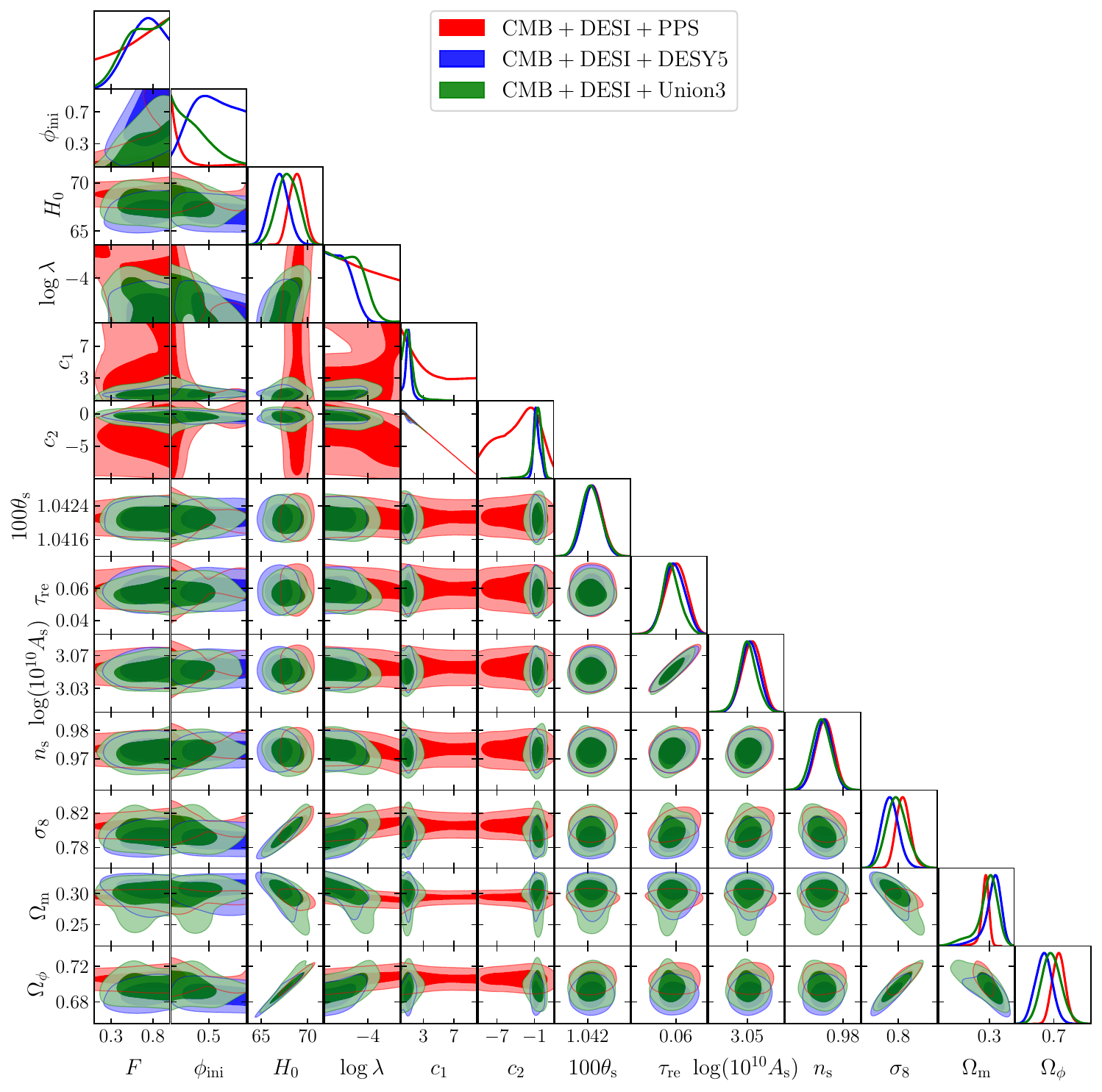}
\caption{The 1D and 2D marginalized posteriors for the $N=2$ case with $c_2\neq c_1$.  }
\label{fig3A}
\end{centering}
\end{figure}

\begin{figure}[H]
\begin{centering}
\includegraphics[width=1.0\linewidth]{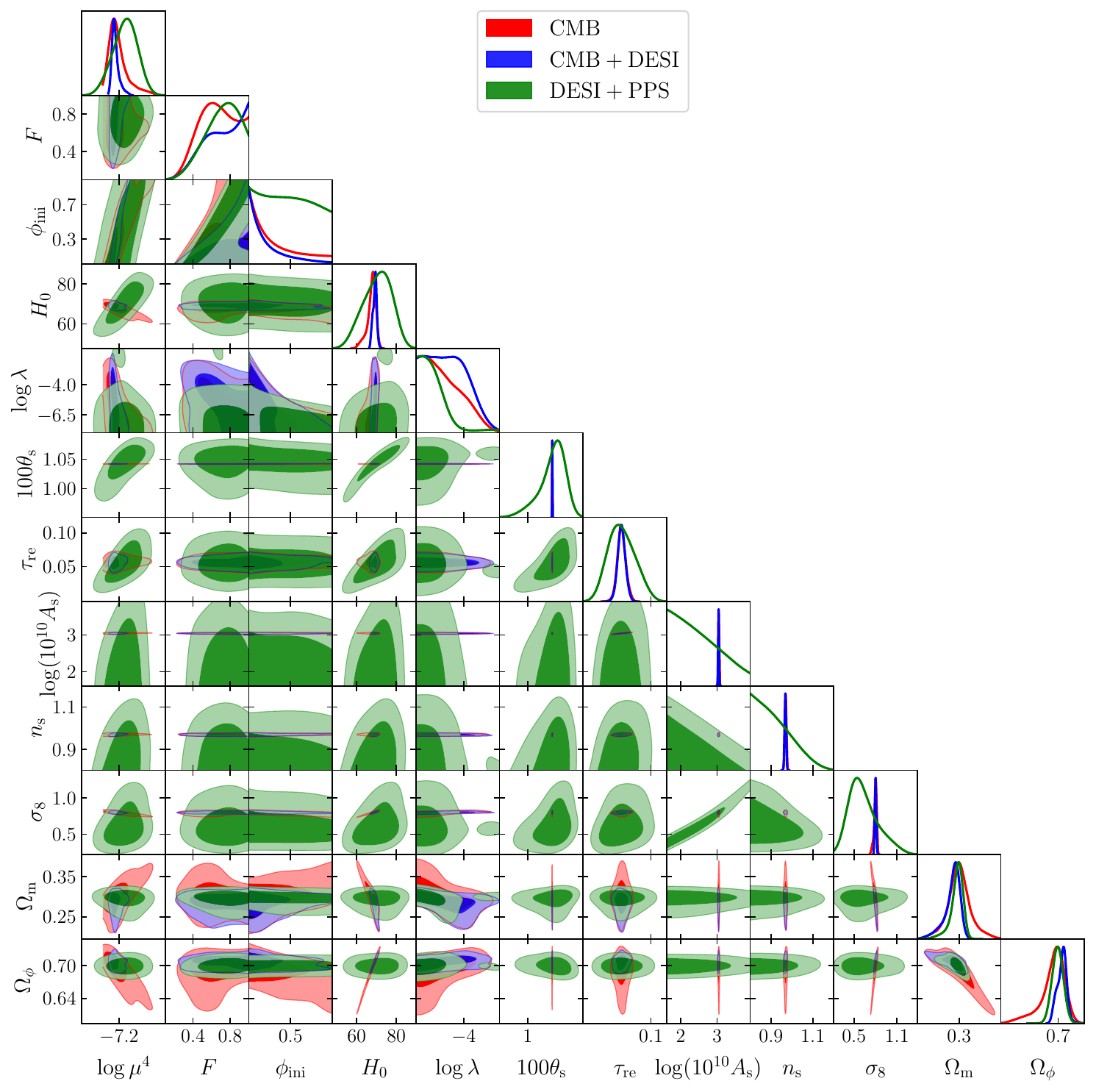}
\caption{The 1D and 2D marginalized posteriors for the $N=1$ case using the CMB-only and CMB+DESI data sets and the late-time only data set DESI+PPS. }
\label{fig4A}
\end{centering}
\end{figure}

\begin{figure}[H]
\begin{centering}
\includegraphics[width=1.0\linewidth]{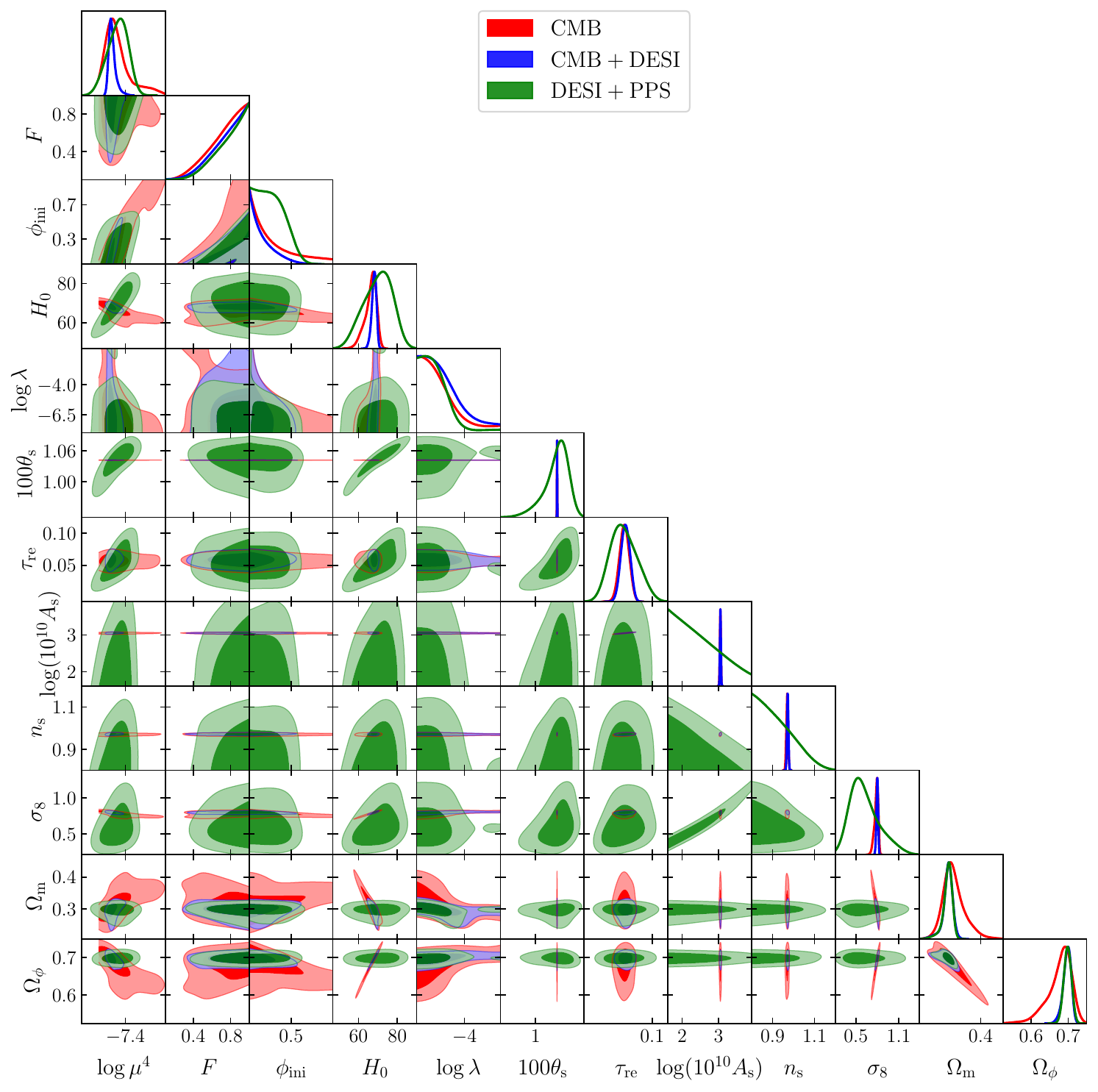}
\caption{The 1D and 2D marginalized posteriors for the $N=2$ case (with $c_2=c_1$) using the CMB-only and CMB+DESI data sets and the late-time only data set DESI+PPS.  }
\label{fig5A}
\end{centering}
\end{figure}

\end{document}